\documentclass[12pt]{article}
\usepackage{amsmath,amssymb,amsthm,color}
\usepackage{graphicx}
\usepackage{cite}
\usepackage{epsfig}
\usepackage{lineno}
\renewcommand{\baselinestretch}{1.66}

\begin{document}

\title {Magneto-electronic properties of twisted bilayer graphene system}
\author{
\small Chiun-Yan Lin$^{a}$, Ming-Fa Lin$^{a,b*}$ $$\\
\small  $^a$Department of Physics, National Cheng Kung University, Tainan 701, Taiwan \\
\small  $^b$Hierarchical Green-Energy Materials/quantum topology centers, \\
\small  National Cheng Kung University, Tainan 701, Taiwan }

\renewcommand{\baselinestretch}{1.66}
\maketitle

\renewcommand{\baselinestretch}{1.66}

\begin{abstract}
The generalized tight-binding model is developed to investigate the magneto-electronic properties in twisted bilayer graphene system.
All the interlayer and intralayer atomic interactions are included in the Moire superlattice.
The twisted bilayer graphene system is a zero-gap semiconductor with double-degenerate Dirac-cone structures, and saddle-point energy dispersions appearing at low energies for cases of small twisting angles.
There exist rich and unique magnetic quantization phenomena, in which many Landau-level subgroups are induced due to specific Moire zone folding through modulating the various stacking angles.
The Landau-level spectrum shows hybridized characteristics associated with the those in monolayer, and AA $\&$ AB stackings. The complex relations among the different sublattices on the same and different graphene layers are explored in detail.
\end{abstract}

\par\noindent  * Corresponding author.
{~ Tel:~ +886-6-275-7575.}\\~{{\it E-mail addresses}: mflin@mail.ncku.edu.tw (M.F. Lin)}

\pagebreak
\renewcommand{\baselinestretch}{2}
\newpage

\vskip 0.6 truecm

\section{Introduction}

The achiral and chiral geometric structures play the critical roles in diversifying the essential properties, so they are one of the main-stream topics in basic researches\cite{4Nature556;43,4Nature354;56}, as well as highly potential applications\cite{4AdvMater22;3723}. A cylindrical carbon nanotube could be regarded as a rolled-up graphitic sheet; therefore, its structure is well characterized by two primitive vectors of monolayer graphene\cite{4Nature354;56}. About twenty years ago, the hexagonal geometries and low-energy electronic properties of 1D single-walled carbon nanotubes are accurately identified from the simultaneous high-resolution measurements of STM and STS\cite{4Nature391;59}. The close relations between the chiral angle [the arrangement of hexagons relative to the tubular axis] and the metallic/semiconducting behavior are thoroughly examined by the delicate experiments, directly verifying the theoretical predictions by the tight-binding model\cite{4JPSJ71;1820,4APL60;2204} and first-principles calculations\cite{4PRB66;115416,4PRL84;2917}. That is, the low-lying $\pi$-electronic states in a carbon nanotube is sampled from those of monolayer graphene through the periodical boundary condition. However, the misorientation of ${2p_z}$ orbitals and their hybridizations with three ${\sigma}$ orbitals, which appear in very small cylindrical surfaces, have very strong effects on the low-energy fundamental properties. The experimental examinations on the conducting or semiconducting behaviors have confirmed the critical mechanisms, the periodical boundary condition and the curvature effects\cite{4Science306;666}. The similar geometric structures come to exist in the both ends of 1D graphene nanoribbons with the open boundary conditions\cite{4PCCP18;7573}. The zigzag, armchair and chiral edge structures are clearly classified under the distinct STM experiments\cite{4Nature391;59}. Also, all of them are confirmed to be semiconductors, in which their energy gaps are inversely proportional to the nanoribbon widths\cite{4Nature391;59}, as predicted by the theoretical calculations\cite{4JPSJ71;1820}. The semiconducting properties are principally determined by the open boundary, the non-uniform bond lengths/hopping integrals near two boundaries, and the zigzag-edge magnetism. \\

The twisted bilayer graphene systems have been successfully generated in experimental laboratories through the various methods, such as, the mechanical exfoliation\cite{4APL93;172108}, the chemical vapor deposition method\cite{4Science312;1191,4PRB75;214109}, and transfer-free method\cite{4ACSNano5;8187}. There exists a relative rotation angle [$\theta$] between two honeycomb lattices, being clearly identified from the high-resolution STM measurements\cite{4Nature6;109,4PRL109;196802,4PRL100;125504}.
This significant parameter represents a specific Moire superlattice corresponding to an unit cell much larger than that [four carbon atoms] of the AA and AB stackings. Since two graphene layers are attracted by the van der Waals interactions, the low-energy essential physical properties are expected to be dominated by the very complicated interlayer hopping integrals of C-${2p_z}$ orbitals in the absence of spin-orbital couplings. Furthermore, more pairs of valence and conduction bands will come to exist under the critical zone-folding effect. The rich and unique electronic structures are confirmed by the ARPES\cite{4NatMat12;887,4PRL109;186807}, e.g., the coexistence of linear and parabolic energy dispersions in symmetry-broken bilayer graphene systems\cite{4NatMat12;887}, and the semiconducting or semi-metallic behaviors\cite{4PRL109;186807}. However. the further experimental examinations are required to identify  more valence subbands in the reduced first Brillouin zone. \\

Both STS\cite{4PRL109;196802,4Nature6;109} and transport experiments\cite{4PRB85;201408} on twisted bilayer graphenes could provide enough information on the low-energy physical properties. The former directly measure the band properties/magneto-electronic ones across the Fermi level and the structures of van Hove singularities [the local maxima $\&$ minima, and the saddle points in the energy-wave-vector space]; that is, they are useful in identifying the complex effects due to the Moire superlatices and the external fields. Up to date, the  experimental results clearly show three kinds of van Hove singularities, the V-shaped structure, shoulder, and prominent symmetric peaks at the negative and positive energies\cite{4PRL109;196802,4Nature6;109}, being very sensitive to the twisted angles. Obviously, such measurements indicate the different low-lying energy dispersions/critical points. The unusual energy bands, being due to the zone-folding effects, will be magnetically quantized to the rich and unique Landau levels. In addition, the STS and optical measurements on the magneto-electronic  density of states are absent. Moreover, the quantum transport experiments display the Hall conductivities in analogy with those of bilayer AB stacking or the superposition of two monolayer graphene systems. Very interesting, certain twisted bilayer graphenes, with magic angles of ${\sim\,1.1^\circ}$ (relative to the bilayer AA stacking)\cite{4Nature556;43}, are identified to become the unusual superconductors under a low critical temperature [${T_c\sim\,1.7}$ K. It might need to develop the modified theoretical frameworks for the unique superconducting phenomena in emergent layered materials. \\

On the theoretical progress of twisted bilayer graphene systems, the first-principles method\cite{4PRL109;196802}, the effective-mass approximation\cite{4PRB81;161405,4PRB87;125414}, the tight-binding model\cite{4PRB85;195458} are available in exploring their essential properties. An optimal geometric structure, with the lowest ground state energy, can only be determined by the numerical VASP calculations\cite{4PRL109;196802}; that is, the complete relation between the total energy and the twisted angle is obtained under such evaluations. These three manners might be suitable/reliable in investigating band structures and density of states. In general, the calculated results are roughly consistent with the ARPES measurements, e.g., the modified linear dispersions and the emergence of parabolic dispersions\cite{4NatMat12;887,4PRL109;186807}. Furthermore, they clearly show that the band-edge states [the critical points in the energy-wave-vector space] create the V-shape form, shoulder and logarithmically divergent peaks, respectively, corresponding to the linear Dirac cone, the local maxima/minima and the saddle points. Such van Hove singularities have been identified from the high-resolution STS measurements [discussed earlier; Ref\cite{4PRL109;196802}]. The main features of electronic properties are directly reflected in optical spectra, such as, the number, frequency, and intensity of prominent absorption structures\cite{4PRB87;205404}. The optical spectroscopies are required to verify the theoretical predictions. However, it will be very difficult to investigate the twist-enriched magnetic quantization, mainly owing to the Moire superlattice with the complex interlayer hopping integrals. Only few studies by the effective-mass approximation and the tight-binding model are conducted on the electronic structure and Hall conductivities, e.g., the fractal energy spectrum\cite{4NanoLett12;3833}, the magnetic-field-dependent butterfly diagram [the Aharonov-Bohm effect\cite{4PRB85;195458}], and quantum plateaus with the normal/irregular heights\cite{4PRB85;201408}. Apparently, one of the studying focuses, the characteristics of magnetic wave functions, are absent up to date, since their spatial distribution symmetries will dominate the magneto-selection  rules in dynamic optical transitions and static scattering events. In short, the systematic investigations on the geometric symmetries,  electronic properties, optical excitations, and magneto-electronic states are critical in providing a full understanding of the geometry-diversified physical phenomena for twisted bilayer graphene systems. Maybe, the further development of the generalized tight-binding model and its direct combination with the other modified theories are the most efficient way, as discussed in the following paragraphs. \\

From the theoretical point of view, the generalized tight-binding model and the gradient approximation are available in fully understanding the essential electronic and optical properties. For twisted bilayer graphene systems, the low-energy electronic structures, van Hove singularities in density of states, optical absorption spectra, and magneto-electronic states, which are greatly diversified by the composite effects due to the zone folding, the twist-dependent interlayer hopping integrals, the stacking symmetries, the gate voltages, and the magnetic fields, will be investigated in detail. As to the low-energy physical properties, the studying focuses cover the semimetallic or semiconducting property across the Fermi level, the linear/parabolic/oscillatory energy dispersions, the gate-voltage-dominated state degeneracy, the creation of more band-edge states, the distinct special forms of  2D van Hove singularities, the twisting-enriched optical absorption spectra [the threshold and higher-frequency excitation structures], the pair/group number of valence and conduction bands/Landau levels, and the $B_z$- and $V_z$-dependent magneto-electronic energy spectra and wave functions with the non-crossing, crossing and anti-crossing behaviors. Also, the concise physical pictures, being associated with to the significant cooperations/competitions among the zone-folding effect, the distinct interlayer hopping integrals, the gate voltages and the magnetic fields, are proposed to account for the diverse physical phenomena. A detailed comparison with the sliding bilayer graphenes is made, clearly illustrating the significant differences in fundamental properties as a result of the distinct stacking symmetries. \\

\section{Electronic properties in the absence/presence of gate voltages}

The specific geometry of a twisted bilayer graphene can be well defined by the primitive lattice vectors of honeycomb lattices, as done for any single-walled carbon nanotubes\cite{4Nature354;56}. It is characterized by the relative rotation angle [$\theta$] and translation vector between two graphitic sheets. The primitive unit cell of the commensurate graphene bilayer structure is clearly shown in Fig. 1, corresponds to the least common multiples of the two distinct graphene layers. The primitive unit vector $G_1$ expressed as
\begin{align}
\bf G_1\,=m\bf{a_1^{(1)}}+n{\bf a_2^{(1)}}=m^{\prime}\bf{ a_1^{(2)}}+n^{\prime}\bf{ a_2^{(2)}}. \tag{1}
\end{align}
Furthermore, the $G_2$ has the same magnitude, but presents a relative rotation angle of ${60^\circ}$.
Both ${\bf a_1}$ and ${\bf a_2}$ are unit vectors of a honeycomb, in which the superscripts represent the first and second layers. Through the specific relation among them\cite{4PRB81;161405}, the identical coefficients are revealed as ${m=n^\prime}$ $\&$ ${n=m^\prime}$. Apparently, a single pair of ${[m, n]}$ is sufficient in describing a twisted  bilayer graphene, leading to the rotation angle of ${{cos(\theta\,)},=(m^2+n^2+4mn)/2(m^2+n^2+mn)}$. In general, this angle is periodical within the range of ${60^\circ\,<\theta\,<\,0^\circ}$, where the minimum and maximum ones are, respectively, the regular AA and AB stackings [the non-twisted systems]. According to a hexagon with two carbons, the total number [$N_M$] in a Moire superlattice is ${4(m^2+n^2+mn)}$; that is, there are $N_M$ ${2p_z}$-orbitals which  dominate the low-energy essential properties, e.g., the band structures, optical excitations and magnetic quantizations. For example, a ${N_M\times\,N_M}$ Hamiltonian matrix, with the intrinsic atomic interactions in Eq. (2) and (3), is sufficient in exploring electronic and optical phenomena under a vanishing magnetic field.

The low-energy tight-binding model, which mainly originates from the intralayer $\&$ interlayer atomic interactions of carbon-2p$_z$ orbitals, is expressed as
\begin{equation}
H=-\sum_{{\bf R_{i}},{\bf R_{i}}}t({\bf R_{i}},{\bf R_{j}})[c^{+}({\bf R_{i}})c({\bf R_{j}})+h.c.], \tag{2}
\end{equation}
where $c^{+}({\bf R_{i}})$ and $c({\bf R_{j}})$ are creation and annihilation operators at positions ${\bf R_{i}}$ and ${\bf R_{j}}$, respectively. The various hopping integrals are characterized under the empirical formula
\begin{equation}
-t({\bf R_{i}},{\bf R_{j}})=\gamma_0 e^{-\frac{d-b}{\rho}}(1-(\frac{d_0}{d})^2) + \gamma_1 e^{-\frac{d-d_0}{\rho}} (\frac{d_0}{d})^2, \tag{3}
\end{equation}
where $d=|{\bf R_{i}}-{\bf R_{j}}|$ is the distance between two lattice points, $b=1.42$ {\AA} the in-plane C-C bond length, $d_0=3.35$ {\AA} the interlayer distance, and $\rho = 0.184 b$ the characteristic decay length. $\gamma_0 = -2.7$ eV is the intralayer nearest-neighbor hopping integral and $\gamma_1 = 0.48$ eV the interlayer vertical atomic interaction. This model is reliable and suitable for investigating the essential properties of the geometry-modulated bilayer graphene. In the presence of an external perpendicular electric field, the opposite on-site energies between two layers is used to simulate the layer-dependent Coulomb potentials. \\

The bilayer graphene systems, including the normal and twisted stacking configurations, exhibit the rich band structures, mainly owing to the regular/enlarged unit cells and distinct interlayer hopping integrals. For example, the AA and AB stackings [${\theta\,=0^\circ}$ and 60$^\circ]$], with only four carbon atoms in a primitive cell, possess two pairs of low-lying valence and conduction bands which are initiated from the K/K$^\prime$ valleys, as clearly displayed in Figs. 2(a) and 2(b). The main features of the former cover two vertically separated Dirac-cone structures in the linear and isotropic form, and free valence holes and conduction electrons due to the AA stacking symmetry. The Dirac-point energies are dominated by the vertical interlayer hopping integral [details in Ref.\cite{4IOP;CY}]. The energy spacing of two Dirac points gradually grows in the increase of gate voltage. This also leads to the significant enhancement of the Fermi-momentum states and thus the 2D free carrier densities. It should be noticed that such zero band-gap band structure will create an optical excitation gap [discussed later in Fig. 6]. As to the higher-/deeper-energy states [$\sim$${|E^{c,v}|\ge\,2}$ eV], the M valley is generated by the saddle point. On the other hand, electronic states of the latter, as shown in Fig 2(b), have parabolic energy dispersions near the K point under a very weak valence and conduction overlap [inset]. Two regular stackings belong to semimetals because of a finite density of states at the Fermi level [Fig 5]. Very interesting, a direct band gap is opened by the gate voltage; that is, the semimetal-semiconductor transition occurs during the variation of ${V_z}$. Furthermore, the monotonous energy dispersions of the first pair are dramatically changed into the oscillatory ones, where the extreme band-edge states consist of the unusual 1D constant-energy loops in the energy-wave-vector space. \\

General speaking, the valence and conduction bands are drastically changed by the twisted angles. At low energies near the Fermi level, the ${\theta\neq\,0^\circ}$ $\&$ ${60^\circ}$ bilayer graphene systems, as clearly indicated in Figs. 2(c)-2(d), 3(a) and 4(a), exhibit the degenerate Dirac-cone structures at the K/K$^\prime$ valleys with the almost isotropic energy spectra. The Fermi velocities are different for the conduction and valence states; furthermore, their magnitudes quickly declines as $\theta$ decrease as a result of more complex interlayer hopping integrals. Such results could be roughly regarded as the direct superposition of those in separate monolayer graphene within the reduced hexagonal first Brillouin zone; that is, the Moire superlattice of bilayer graphenes creates the non-uniform interlayer hopping integrals, while it hardly affects the low-energy fundamental properties, being quite different from the AA and AB stackings [Figs. 2(a) and 2(b)]. All the twisted materials only belong to zero-gap semiconductors, in analogy with a single-layer graphene. Each $\theta$-dependent system  presents ${N_M/2}$ pairs of valence and conduction subbands within the  whole energy range, being directly reflected in the density of states, optical absorption spectra and magneto-electronic properties. \\

Apparently, the significant energy ranges, which correspond to the different valleys, are very sensitive to the variation of twisted angles.  In addition to the lowest K-point valence /conduction valleys including the Fermi level, the first M-point valleys come to exist in the lower-/shallower-energy region for the conduction/valence bands, compared to the AA and AB stackings [Figs. 2(a) and 2(b)].  Furthermore, their state energies quickly grow in the increment of $\theta$, such as, [0.15 eV, ${-0.18}$ eV], [0.42 eV, ${0.45}$ eV], [0.52 eV, ${0.60}$ eV] and [1.60 eV, ${-1.65}$ eV], respectively, originating from ${\theta\,=3.89^\circ}$, ${\theta\,=6.00^\circ}$ ${\theta\,=9.40^\circ}$, and.${\theta\,=21.8^\circ}$ [Figs. 2(c), 2(d), 3(a), and 4(a)].  Very interesting, there exist four conduction/valence subbands associated with the M-point valleys along M$\Gamma$ [marked by the green circles], and they are doubly degenerate under the direction of MK. Such energy dispersions might have the rich saddle points and local extreme ones, being situated at the M point, or close to it along M$\Gamma$. However, few of them do not present any band-edge states. With the increase of state energy, several $\Gamma$-related valleys, which includes those along $\Gamma$K, are stably formed, in which their observations are relatively easy for the smaller-$\theta$ bilayer graphene systems, e.g., ${\theta\,=3.89^\circ}$ in Fig. 2(c). In addition, the extra few valleys are built form the K point. It should be noticed that both $\theta$- and 60$^\circ$${-\theta}$-dependent  systems present the almost identical band structures, such as, electronic energy spectra of the (1,2) and (1.4) bilayer structures [the solid and dashed curves in Fig. 4(a)]. Apparently, the band-edge states in the different valleys are expected to exhibit the unusual van Hove singularities and initiate the highly degenerate Landau levels, and their diversified features are due to the combined effects of the zone folding and the complex stacking symmetry. \\

Electronic structures of twisted bilayer graphene systems are greatly enriched by the external gate voltages, especially for those of the Dirac-cone valleys initiated from the K/K$^\prime$ points. The double degeneracy, as shown in Figs. 3(a)-3(d) and 4(a)-4(d), is obviously broken in the presence of layer-dependent Coulomb potential energies. The non-equivalence of the top and bottom honeycomb lattices is principally responsible for the low-energy state splitting. The Fermi-momentum states gradually deviate from the K point in the increase of gate voltage. At zero ${V_z}$, the free carriers fully disappear as a result of the absence of valence and conduction band overlap. However, a uniform perpendicular electric field has successfully generated the vertical separation of two isotropic Dirac-cone structures, clearly illustrating their significant overlap across the Fermi level. This result indicates that both valence holes and conduction electrons, respectively, corresponding to the upper and lower Dirac cones, come to exist simultaneously. Furthermore, the 2D free carrier densities grow with the increasing/decreasing of gate voltages/twisted angles. Very interesting, the cooperation of complex interlayer interactions and layer-dependent Coulomb potentials lead to the split K and K$^\prime$ valleys, such as the different Fermi momenta for free electrons and holes in the insets of Figs  4(c)-4(d). This result  clearly illustrates the critical role of the latter in breaking the equivalence of ${[A^1, B^1]}$ sublattices. It is further expected to crate the valley-related Landau levels. In addition, the splitting of electronic states might appear in other valleys. The above-mentioned features of energy bands will be directly reflected in more complicated van Hove singularities and optical absorption structures. \\

The angle-resolved photoemission spectroscopy [ARPES], which possesses the very high resolutions in the measurements of energies and momenta, is the only tool in identifying the wave-vector-dependent quasi-particle energy spectra and widths for the occupied electronic states [details in Refs \cite{4NatMat12;887,4PRL109;186807}]. The measured energy  dispersion relations can directly examine the theoretical calculations from the tight-binding model and the first-principles method. In general, the ARPES chamber is accompanied with the instrument of sample synthesis to measure the ${\it in}$-${\it situ}$ band structures. Up to now. the experimental measurements have confirmed the feature-rich occupied band structures in the graphene-related systems with sp$^2$ bondings, as observed under the various dimensions\cite{4ACSNano6;6930,4PRB73;045124}, layer numbers\cite{4NanoLett17;1564}, stacking configurations\cite{4NanoLett17;1564}, substrates\cite{4PRB78;201408}, and adatom/molecule chemisorptions\cite{4PRB83;125428}. The rich and unique electronic energy spectra cover the 1D parabolic energy subbands with the specific energy spacings and band gap in graphene nanoribbons\cite{4ACSNano6;6930}, the downward valence Dirac cone in monolayer graphene\cite{4PRB78;201408}, two pairs of 2D parabolic bands in bilayer AB stacking\cite{4NatMat12;887} the coexisting linear and parabolic dispersions in symmetry-broken bilayer graphenes\cite{4NatMat12;887}, the linear and parabolic bands in tri-layer ABA stacking\cite{4NanoLett17;1564}, the linear, partially flat and sombrero-shaped bands in tri-layer ABC stacking\cite{4NanoLett17;1564}, the substrate-induced large energy spacing between the $\pi$ and $\pi^\ast$ bands in bilayer AB stacking \cite{Zhou;770}, the substrate-induced oscillatory bands in few-layer ABC stacking\cite{4PRB78;201408}, the semimetal-semiconductor transitions and the tunable low-lying energy bands after the molecule/adatom adsorptions on graphene surface\cite{4PRB83;125428}, the 3D band structure, with the bilayer- and monolayer-like energy dispersions, respectively, at ${k_z=0}$ = 0 and 1 [K and H points in the 3D first Brillouin zone] and the strong wrapping effect along the KH axis, for the AB-stacked [Bernal] graphite\cite{4PRB73;045124}. Specifically, the theoretical predictions on the low-lying valence bands of the twisted bilayer graphene systems are verified by the delicate ARPES analysis and examinations, such as the valence Dirac-cone structure of the K-point valley and the parabolic dispersions near the saddle M-point\cite{4PRB85;195458}. The other stable valleys [Fig. 2], which are associated with the band-edges states at the deeper energies, are worthy of the further verifications.
Furthermore, the gate-voltage-enriched energy dispersions. the $V_z$-induced splitting of K and K$^\prime$ valleys, and semiconductor-semimetal transitions require the experimental checks. They can provide the full information on the complicated interlayer hopping integrals.

The main features of energy bands in bilayer graphene systems result in the rich van Hove singularities. Their densities of states, which are the C-${2p_z}$ orbitals, as shown in Figs. 5(a)-5(f), sharply contrast between the normal and twisted stackings. The AA bilayer stacking [the black curve in Fig. 5(a)] has one plateau and a pair of cusp structures across the Fermi level, directly reflecting the semimetallic behavior of two separated vertical Dirac cones near the K/K$^\prime$ point [Fig. 2(a)]. Furthermore, another pair of prominent peaks, which are divergent in the logarithmic form, originate from the saddle-M-point valleys at ${\sim\,2.00}$ eV $\&$ 2.80 eV/${\sim\,-2.50}$ eV $\&$ ${-3.50}$ eV for conduction-/valence-band states. But for the AB-stacked system [Fig. 5(b)], only a pair of neighboring shoulders, with a narrow energy spacing [${\sim\,20\,}$ meV in the inset], appear at the left- and right-hand sides of ${E_F=0}$. It also belongs to a semimetal, but presents the lower free carrier density compared with the AA case. At higher/deeper energies, the energy spacings of strong symmetric peaks become smaller, and even vanishing under the merged conduction valley [Fig. 2(b)]. Apparently, the special structures of the AA stacking are somewhat widened and lowered during the increment of gate voltage [the red and blue curves]. On the other hand, the AB stacking exhibits the dramatic transformation in the low-lying structures and the obvious splitting of prominent conduction peaks. That is to say, a conduction/valence shoulder structure is replaced by an asymmetric peak in the square-root divergent form and a shoulder with an observable band gap. Specifically, such peaks are created by the extreme constant-energy-loops, since their ${[k_x, k_y]}$-states consist of the effective 1D parabolic energy dispersions. \\

Because of the ${\theta\neq\,0^\circ}$ twisting effects, these bilayer graphenes exhibit the unique characteristics in the density of states, as clearly indicated by Figs. 5(c)-5(f). All the twisted systems have a V-shape structure with a zero DOS at the Fermi level and the different slopes in the left- and right-hand sides. This result directly reflects the degenerate Dirac cones in the distinct Fermi velocities [inversely proportional to DOS; Figs. 2(c), 2(d), 3(a) and 4(a)]; therefore, the vanishing valence and conduction band overlaps dominate the zero-gap semiconducting behaviors. When state energies become higher/deeper, a prominent symmetric peak in the logarithmic form comes to exist in the positive and negative regions, being followed by the obvious shoulders.  Such van Hove singularities mainly originate from the saddle and extreme band-edge states near the M point. Furthermore, their energies are very sensitive to the twisted angles. For example, the separated/composite special structures are more close to the Fermi level for a  smaller $\theta$. Apparently, they also appear at the other range ranges as a result of the saddle M-point valleys and the extreme M-, $\Gamma$- and K-point ones. Specifically, the gate voltages create the splitting of band-edge states and thus might induce some split van Hove singularities. The specific unusual structure, which crosses the Fermi level, is revealed as a plateau and two-side cusp structures. It represents a semimetallic system in the presence of a significant overlap behind two vertical Dirac-cone energy spectra/two pairs of valence and conduction bands. \\

The high-resolution STS measurements, which can clearly identify the van Hove singularities due to the low-lying critical points of valence and conduction states, are very useful in examining the relation between  stacking configurations and energy bands in the twisted bilayer graphene systems. Up to now, the experimental examinations are consistent with the theoretical predictions [Fig. 5 ; Refs\cite{4PRL109;196802,4Nature6;109}]. As to the various $\theta$-dependent bilayer materials, they have verified a pair symmetric peaks in the logarithmic form below and above the Fermi level, obviously arising from the lower-energy saddle M-point  [Figs. 2 and 4]. For example, the energy spacings between the valence and conduction peak structures are observed to be 12 meV, 82 meV and 430 meV for ${\theta\,=1.16^\circ}$, 1.79$^\circ$ and 3.40$^\circ$, respectively\cite{4Nature6;109}. Apparently, the saddle-point energies are very sensitive to the twisted angles, as further confirmed by mre cases \cite{4PRL109;196802}. From the theoretical and experimental results, the very large  Moire superlattices, with the significant zone-folding effects,  are capable of greatly reduce the saddle-point energies, such as the obvious difference between the twisted bilayer and monolayer graphene [${\sim\,\pm\,2}$ eV] systems. \\

\section{Optical absorption spectra}

The rich electronic energy spectra and wave functions are responsible for the unusual optical properties In the twisted bilayer graphene systems. The vertical photon excitations are the transition channels from the occupied states to the unoccupied ones with the same wave vectors; therefore, they strongly depends on the joint density of states (JDOS) related to the initial and final states and the dipole matrix elements. The former, JDOS, might be dominated by the van Hove singularities of the band-edge states; that is, the frequency, form and intensity of absorption structures would be determined by them. Furthermore, the wave functions of such critical points could create the finite dipole moments and thus the available optical excitations due to the almost symmetric valence$\rightarrow$conduction channels. The typical stacking configurations are chosen for a model study, and the significant differences among the $\theta$-dependent bilayer materials are explored in detail. As discussed earlier in the $V_z$-diversified band structures [Figs. 3 and 4], the gate voltage is expected to induce the dramatic transformation in absorption spectra, especial for the threshold structure associated with the semiconductor-semimetal transition. \\

The optical excitation spectra of graphene systems exhibit the diverse absorption structures through the twisting-induced stacking configurations. A single-layer graphene, as shown in Fig. 6(a) by the dashed black curve, exhibits the linear $\omega$-dependence at the low absorption frequency. An optical gap, the threshold frequency, is vanishing; furthermore, absorption structures are absent below [${2\gamma}$ $\sim\,5$ eV]. These results directly reflect energy spectrum and wave functions of the valence and conduction Dirac cones. It should be noticed that the low-frequency dipole matrix element is proportional to the Fermi velocity [${v_F=3\gamma_0\,b/2}$]\cite{4IOP;CY}. In addition, there exists a strong $\pi$-electronic absorption peak at ${\omega\,=2\gamma_0}$, being initiated from the saddle M point. Such prominent symmetric structure is frequently observed in the carbon-sp$^2$ bonding systems\cite{4IOP;CY}. \\

The AA-stacked bilayer graphene. with ${\theta\,=0^\circ}$, presents the unusual low-frequency absorption spectra, as indicated in Fig. 6(a) by the solid black curve. The critical picture is the well-behaved two pairs of valence and conduction Dirac cones almost symmetric about the Fermi level [Fig. 2(a)]. Their wave functions are the symmetric or antisymmetric linear superposition of the layer- and sublattice-dependent four tight-binding functions\cite{4IOP;CY}. As a result, the available excitation channels only originate from the same Dirac-cone structure, in which their diploe moments are roughly same with that of monolayer graphene\cite{4IOP;CY}. That is to say, the inter-Dirac-cone vertical transitions are forbidden during the vertical optical excitations, mainly owing to vanishing dipole matrix elements. The first and second Dirac-cone structures, respectively, possess the band-overlap-induced free holes and electrons; therefore, the Fermi-Dirac distribution functions will dominate the threshold absorption frequency and structure. From the analytic energy spectrum of the bilayer AA stacking\cite{4JAP110;013725}, its optical gap, being accompanied with the initial shoulder structure, is examined to be approximately double of that the vertical interlayer hopping integral [${\sim\,2\gamma_1}$; \cite{4JAP110;013725}]. The absorption gap is gradually enhanced by the external gate voltages, The comparable absorption frequencies are predicted for the AA-stacked graphenes with the even layer numbers. However, the $N$-odd few-layer systems have the zero threshold frequency, since the Dirac point of the middle cone structure touches with the Fermi level. For example, the trilayer material presents the composite absorption spectra of monolayer and bilayer ones. Apparently, the low- and middle-frequency optical citations are, respectively, dominated by the electronic states close to the K and M points. \\

The ${\theta\,=60^\circ}$ bilayer AB stacking, as illustrated in Fig. 6(b), clearly displays the unique absorption spectra. According to band structure in Fig. 2(b), the vertical threshold excitations are closely related to the first pair of parabolic valence and conduction bands across the Fermi level. Their dipole matrix elements near the band-edge states are finite, so that both optical and band gaps are zero. Specifically, those, which are due to the second pair of parabolic energy dispersions, do not create the finite contribution, i.e., the absence of absorption shoulder structure at ${\omega\,\sim\,2\gamma_1}$. However, there is a broadened discontinuity, being revealed at ${\omega\,\gamma_1}$. This specific shoulder structure arises from the band-edge-state excitations of the first/second valence band and the second/first conduction one. In addition, an optical gap is presented when the external gate voltage is sufficiently large\cite{4IOP;CY}. \\

The ${\theta\neq\,0^\circ}$ and 6$^\circ$ twisted bilayer graphene systems display a lot of absorption structures below the middle-frequency $\pi$-electronic absorption peak [$\omega < 5$ eV], as indicated in Figs. 6(c), 6(d) and 7(a) by the solid black curves. The lower-frequency absorption spectra, which are due to the degenerate valence and conduction Dirac cones of the K/K$^\prime$ valley [Figs. 2(c), 2(d) and 4], reveals the linear $\omega$-dependence in the spectral intensity. The similar result is presented in monolayer graphene [the dashed curve in Fig. 6(a)] because of the almost identical dipole moments. With the increasing absorption frequency, three observable/four prominent symmetric peaks come to exist as a result of the electronic states near the saddle M point [Fig. 7(b)]. For example, the (1,2) bilayer graphene exhibit the specific absorption peaks at ${\omega_a\sim\,}$3.11 eV, 3.35 eV  and 3.45 eV [Fig. 7(a)]. By the delicate analysis using the absorption spectrum and joint density of states [the red curve], the first, second and third peaks originate from the vertical transitions of the band-edge states near saddle M-point. respectively, corresponding to  [the shallower valence band, the lower conduction one].  [the shallower/deeper valence band, the higher/lower conduction one] and  [the deeper valence band, the higher conduction one], as clearly indicated by the dashed blue lines in Fig. 7(b). That is, all the vertical excitations are available in the first and second pairs of parabolic valence and conduction bands. . The absorption frequency of the initial prominent peaks is obviously reduced through the decrease of the twisted angle, e.g., ${\omega_a\,0.62}$ eV and ${0.28}$ eV for the bilayer (5,6) and (8,9) bilayer graphene systems, respectively [Fig. 6(c) and 6(d)]. Apparently, the frequency, intensity and number of the  $\pi$-electronic absorption peaks in bilayer graphene materials are very sensitive to the change of twisted angle, especially for the wide-range variation of absorption frequency [${\omega_a\,\sim\,0.1-6.0}$ eV.  Generally speaking, it is relatively easy to observe the initial strong  peaks in the small-$\theta$ bilayer systems from the high-resolution optical measurements. Also, there exist other obvious absorption peaks and shoulders at higher excitation frequencies, such as those in the (1,2) bilayer system indicated by the dashed purple lines in Fig. 7(b). Such absorption structures are clearly identified from the critical points near the K, M and $\gamma$ valleys. They become more complicated for the twisted bilayer graphenes with smaller $\theta$'s [Figs. 6(c) and 6(d)]. \\

The vertical optical excitations are greatly enriched by the external gate voltages through the strong cooperations/competitions between the layer-dependent hopping integrals and Coulomb potential energies, e.g., the ${V_z}$-modified absorption spectra of the ${(1,2)}$ bilayer system. The optical gap of the AA bilayer stacking is getting large  in the increment of ${V_z}$ [the dashed red curve in Fig. 9], mainly owing to the enlarged energy spacing of two vertical Dirac points. As for the bilayer AB stacking, the threshold absorption frequency is required to match with the opening of energy gap [the dashed blue curve in Fig. 9]; that is, it is associated with the specific semimetal-semiconductor transition, as identified in the experimental transport measurements \cite{Li;322}. Very interesting, the unusual semiconductor-semimetal transition in any twisted bilayer systems [e. g., the (1,2) system in Fig. 4], creates an optical gap even in the presence of gate voltages, as clearly illustrated in Figs. 8(a)-8(d). The destruction of the double degeneracy in valence and conduction Dirac-cones, which s to two vertical Dirac ones, is responsible this dramatic transformation. The similar result is revealed in the pristine AA case [Fig. 6(a)]. This threshold absorption frequency grows with the increasing gate voltage [Fig. 9]. Moreover, the other/extra absorption structures are modified/created during the variation of $V_z$. For example, the (1,2) bilayer graphene exhibits an extra absorption peak, being due to the band-edge states in the first pair of valence and conduction bands near the M point [Fig. 8(d)], might appear and become an initial one, such as, the ${\omega_a=2.95}$-eV absorption peak at ${V_z=0.1}$ eV [Fig. 8(c)]. Also, the gate-voltage dependence of its excitation frequencies due to the initial four absorption structures is very important, since the calculated could provide the critical $V_z$'s in observing the emergent transition channels. \\

The optical absorption\cite{4Science320;1308}, transmission \cite{4PRL101;267601}, and reflection \cite{4Science320;206} spectroscopies are available in examining the theoretical predictions on the vertical transitions. Up to date, the high-resolution measurements on bilayer graphene systems have confirmed the rich absorption structures only for the AB stacking, covering the ${\sim\,0.3}$-eV shoulder structure under a vanishing field\cite{4PRL101;267601}, the semimetal-semiconductor transition and two low-frequency asymmetric peaks for the sufficiently large gate voltages, and two rather strong $\pi$-electronic absorption peaks at middle frequency\cite{4PRL101;267601}. The  delicate optical experiments are required to verify the theoretical predictions for the ${\theta\neq\,0^\circ}$ and 60$^\circ$ twisted bilayer graphenes, e.g., the zero threshold frequency for any systems, the initial prominent absorption peaks strongly depending on $\theta$, the other higher-frequency excitation structures, the gate-voltage-created optical gaps similar to the AA case, and the $V_z$-enriched/modified transition channels. Such optical ions, as done by the ARPES and STS measurements, clearly identify the composite effects due to the zone folding and complex interlayer hopping integrals in the Moire superlattices. \\

\section{Magnetically quantized Landau levels from the Moire superlattice}

The generalized tight-binding model is reliable in characterizing the magnetic wave functions, while the opposite is true for the effective-mass approximation. By the delicate calculations, one can derive the analytic formulas for the independent matrix elements of the magnetic Hamiltonian. The moire superlattices are very different for the twisted bilayer graphene systems; therefore, it is impossible to get the $\theta$-dependent equations. Only the specific cases, which are full of the important features in magnetic quantizations, are suitable for the model studied. For example, the (1,2)-twisted bilayer graphene has the smallest number of the tight-binding functions, with the fourteen [${A^l_i, B^l_i}$] sublattices in each layer [Fig. 1(b)]. The new spatial period, which is due to the magnetic Peierls phase, is commensurate to that of the Moire superlattice, e.g., the dimension of $\sim3000\times 3000$ for the Hamiltonian Hermitian matrix at ${B_z=100}$ T. Furthermore, the non-vanishing matrix elements are complex numbers even for the ${[k_x=0, k_y=0]}$ state, so the numerical calculations are more complicated in the twisted systems, but not under the normal stackings [AA, AB, ABC and AAB configurations; \cite{4IOP;CY}]. It should be noticed that the Coulomb gauge ${\bf {A}=\frac{\sqrt{3}B_{z}}{6n_{0}}}\times[(2m+n)\widehat{x}-\sqrt{3}n\widehat{y}]
\times[(n-m)\widehat{x}+\sqrt{3}(m+n)\widehat{y}$ is along $G_{1}$ and utilized for the twisted bilayer graphene. \\

First, the main features of magnetic quantizations are presented for monolayer graphene and bilayer  AA $\&$ AB stackings. With the chosen gauge of ${\bf A\,=[-B_{z}y, 0, 0]}$, the magnetic-field-modified unit cell is an enlarged unit cell along the armchair direction, as revealed in Fig. 2. For monolayer systems [AA and AB configurations],  there are 4$R_B$ carbon atoms/C-${2p_z}$ tight-bindig functions [8$R_B$ carbon atoms], being classified into both A and B sublattices [(A$^1$, B$^1$) $\&$(A$^2$, B$^2$)]. Their low-energy magneto-electronic states are directly quantized from the upward conduction K/K$^\prime$ valleys and the downward valence ones [Figs. 2(a) and 2(b)]. Each ${(k_x, k_y)}$ state has the eight-fold degeneracy in terms of four localization centers [related to ${\pm\,B_z\hat z}$ and K $\&$ K$^\prime$ valleys] and spin degree of freedom. Under ${k_x=0}$ and ${k_y-0}$, the oscillatory Landau-level wave functions are localized at the 1/6, 2.6, 4/6 and 5/6 of the enlarged unit cell, e.g., the first case in Figs. 10(a)-10(c). Apparently, the dominating sublattice in monolayer graphene belongs to the A/B sublattice for the 1/6/2/6 localization center; that is, the quantum number of Landau level [$n^{c,v}$] is characterized by its zero-point number [Fig. 10(a)]. The significant difference of zero points in A and B sublattices is equal to one, indicting the full equivalence between them. This property remains unchanged the normal AA and AB stackings because of the identical carbon-atom honeycomb lattices. When the interlayer chemical $\&$ physical environments are taken into account, the oscillation modes of four sublattices are quite different in these two systems. For the former, both  [A$^1$, A$^2$] and [B$^1$, B$^2$] exhibit the totally same oscillations [Fig. 10(b)]. However, the latter [Fig. 10(c)]  presents only the similar oscillations on [A$^1$, A$^2$] sublattices and the mode difference of 2 on [B$^1$, B$^2$] sublattices with the distinct C-atom projections. Concerning the magneto-electronic energy spectra, [monolayer, bilayer AA] and AB, respectively, display the massless and massive behaviors [the ${\sqrt {n^{c,v}B_z}}$ and ${n^{c,v}B_z}$ dependences. IN addition, there are two groups of valence and conduction Landau levels associated with the stable K/K$^\prime$ valleys. \\

The (1,2) bilayer graphene is very suitable for a model study on the rich magnetic magnetization due to the Moire superlattice. Apparently, this system will present fourteen subgroups of conduction and valence Landau level, as magnetically quantized from the zero-field band structure [Fig. 4(a)]. Here, only the two subgroups, which are closely related to the low-energy Dirac cones arising from  from the K and K$^\prime$ valleys, are fully explored for their main features [Figs. 11 and 14]. After diagonalizing the exact Hamiltonian matrix with a lot of complex elements, the Landau-level wave functions and the  $B_z$-dependent energy spectra are obtained through the reliable and efficient way. In addition, an approximate Hamiltonian is not suitable for  investigating the magneto-electronic properties. Very interesting, there exists the eight-fold degeneracy in each Landau, covering the same first and the second subgroups, the double-degenerate composite localization centers [due to ${\pm\,B_z\hat z}$], and the independent spin-up and spin-down configurations [ignored later]. For example, at ${B_z=100}$ T [the second red point in Fig. 14], the second conduction Landau levels exhibit the unusual probability distributions in the magnetic-field-enlarged unit cell, as clearly indicated by [$\alpha$, $\beta$, $\gamma$, $\delta$] in Fig. 11 for the four-fold degenerate Landau states without spin arrangements. The oscillatory magnetic wave functions, which consist of fourteen subenvelope functions in each graphene layer, are localized about [2/6, 4/6] simultaneously on the upper/lower layer [the black/red curves], and near 2/6 or 4/6 singly under the opposite cases. That is to  say, the $\alpha$ ($\beta$) state, respectively, exhibits [2/6, 4/6] and 2/6 (4/6 and [2/6 4/6]) localization modes on the upper and lower layers. Furthermore, the interchange of 2/6 and 4/6 localization centers in the $\alpha$ and $\beta$ degenerate states just corresponds to the $\gamma$  and $\delta$ ones. The critical factors in creating the four-fold degenerate Landau levels come from the equivalence of two layers and localization centers [the linear superposition of the tight-binding functions related to them]. \\

The main features of magnetic oscillation modes in the twisted (1,2) bilayer graphene deserve a closer examination, especially for the unusual relations with those due to monolayer, and bilayer AA $\&$ AB stackings. From the viewpoint of 7 ${A^l_i}$/7 ${B^l_i}$ sublattices on two single-layer graphenes [Fig. 12(a)], the non-uniform physical/chemical environment, which creates the complex interlayer hopping integrals, is response for the modified/non-identical probability distributions, as shown in Fig. 11. For example, this result is clearly illustrated by the simultaneous localization of [2/6, 4/6] on the upper/lower layer of the $\alpha$-/$\beta$-state Landau level. Generally speaking, all the magnetic wave functions do not exhibit to the symmetric or anti-symmetric well-behaved spatial distributions; that is, they should belong to the perturbed type with the major [$n$] and minor [${n\pm\,1}$] modes. The dominating oscillation modes of the $\alpha$-state Landau level, being associated with two localization centers, are summarized on the fourteen lattice sites of the upper/lower graphene layer [Fig. 12(b)]. They present three specific relations. First, the lower layer, as shown by the red dots,] only has the 2/6 localization probability distributions with the $n$ and ${n-1}$ modes for the ${A^2_i}$ and ${B^2_i}$ sublattices, respectively. The equivalent results are revealed in the dominating modes of the 4/6 localization on the upper layer [the black dots]. This property behaves like that of monolayer graphene [Fig. 10(a)]. While the planar projection displays a short distance [<b/3], the two lattice sites on two distinct layers possess the same oscillation mode about the 2/6 localization center, as revealed in bilayer AA stacking [Fig. 10(b)]. Specifically, the 10th carbon atom on the upper layer [the balack dot in Fig. 12(a) is just projected into the center of the lower hexagon; therefore, its ${n+1}$ mode about the 2/6; localization center is accompanied with the $n$ and ${n-1}$ modes on another layer [Fig. 1(b)]. Such behavior corresponding to the bilayer AB stacking [Fig. 10(c)]. The similar phenomena are obtained for the $\beta$-, $\gamma$- and $\delta$-state Landau levels. In short, the rich magnetic quantizations in the $\theta$-dependent twisted bilayer graphene systems originate from the highly hybridized characteristics of monolayer system, and AA $\&$ AB stackings. \\

A perpendicular Coulomb potential leads to the layer-dependent site energies and thus the obvious splitting of Dirac-cone structures/the separated Landau-level energy spectra, as clearly illustrated in Figs. 4 and 13. As for the K-/K$^\prime$-induced   magneto-electronic states, the four-fold degeneracy dramatically changes into the doubly one That is to say, the [$\alpha$, $\beta$]/[$\gamma$. $\delta$] states have the same quantized energy under the specific symmetry of ${\pm\,B_z\hat z}$, where the former/the latter display the higher/lower one. The magnetic wave functions are only localized about 2/6 or 4/6 of the $B_z$-enlarged unit cell [Fig. 13]. The coexistence of two localization centers on the upper or lower layer [Fig. 11] is absent through the full breaking of the mirror symmetry about the ${z=0}$ plane. This phenomenon is examined in detail and should be independent of the magnitude of gate voltage [not shown]. Roughly speaking, all the subenvelope functions on the twisting-induced multi-sublattices present the major $n^{c.v}$,   ${n^{c.v}+1}$, or ${n^{c.v}1}$, or they belong to the hybridized states of three oscillation modes. Furthermore, the robust relations on the intralayer $\&$ interlayer neighboring A${^l_i}$ and B${^{l^\prime}_j}$ subalttices could also be understood from those of monolayer; bilayer AA and AB stackings [Fig. 12(c)], as done for the ${V_z=0}$ case [Fig. 12(b)]. It is also noticed that few lattice positions, with the AB stacking configuration, will lead to the destruction of the ${\pm\,1}$ mode difference for the nearest sublattices on the same layer, e.g., those of the 10 $\&$ 9 sites [the first row in Fig. 13;  the black dots in Figs. 12(a) and 12(c)]. Apparently, the equivalence of A${^l}$ and B${^l}$ sublattices is strongly modified by the gate voltage, or the competition among three kinds of geometric symmetries is enhanced by $V_z$. \\

The magnetic-field-dependent electronic energy spectra, as clearly illustrated in Figs. 14 and 15, are closely related to the main features of band structure [Figs. 2-4]. The low-lying Landau levels, being induced by the stable K/K$^\prime$ valleys, exhibit the mono-layer-like behavior at a zero gate voltage. For example, the non-crossing Landau level energies of the (1,2) bilayer graphene are roughly proportional to  [${\sqrt {n^{c,v}B_z}}$] in the wide energy range ${|E^{c,v}|<1.20}$ eV [Fig. 14(a) at $V_z$=0]. Such range is very sensitive to twisted angles and will quickly decline in the decrease of $\theta$. The gate voltage creates two vertical Dirac-cone structures and thus the composite energy spectra of two monolayer-like ones with the frequent crossing phenomena, e.g., Fig. 15(a) at ${V_z=0.34}$ eV, as observed in the bilayer AA case. Very interesting, the highly degenerate Landau levels shows the delta-function-like peaks in density of states, in which their energies/energy spacings, intensities and numbers strongly depend on the magnetic fields [Fig. 14(b)] and gate voltages [Fig. 15(b)]. For the deeper/higher valence/conduction states, the magnetic quantizations are expected to become very complicated as a result of the non-monotonous dispersion relations and the multi-constant energy loops, e.g., the magneto-electronic states associated with the saddle M points. Whether the anti-crossing behaviors come to exist is under the current investigations. The theoretical predictions could be verified from the high-resolution spin-polarized STS. \\

The main features of magneto-electronic energy spectra and wave functions are reflected in density of states, absorption spectra, Hall transport conductivities, and magnetoplasmon modes, where  the essential properties are worthy of the further theoretical calculations. For example, the frequent Landau-level crossings and/or anti-crossings might be created through the unusual dispersion relations near the saddle M points [e.g., Figs. 2(c) and 2(d)]. Obviously, this will lead to the abnormal delta-function-like van Hove singularities in terms of the magnetic-field-dependent energies and numbers\cite{4IOP;CY}, the coexistence of the regular and extra magneto-optical selection rules\cite{4IOP;CY}, the integer and non-integer quantum plateaus\cite{4PCCP19;2952}; the inter-Landau-level single-particle $\&$ collective excitations and the 2D electron-gas-like magnetoplasmons. Except for the first term, how to solve the technical barriers due to the evaluations of the other essential properties is under the current investigations. On the experimental side, only the quantum transport measurements are conducted on the twisted bilayer graphenes\cite{4PRB85;201408}, and the Hall conductivities are the superposition of these in two monolayer systems. This result is consistent with the quantized Landau levels from two completely degenerate Dirac cones crossing the Fermi level [Figs. 4, 11 and 14]. \\

\section{Concluding remarks}

There exist significant differences between the twisted and sliding bilayer graphenes in the fundamental physical properties. The former and the latter are, respectively, characterized by the specific twistings with two commensurate honeycomb lattices, and the relatively displacement of two graphene layers along the armchair direction and then the zigzag one [Fig. 16]: AA$\rightarrow$AB$\rightarrow$AA$^\prime$ $\&$ AA$^\prime$$\rightarrow$AA. As a result, their primitive unit cells are mainly determined by the greatly enlarged Moire superlattice and the AA-like ones, respectively, including many and four carbon atoms. The numerical calculations are relatively easy for the sliding-induced electronic, optical and transport properties, as well as their magnetic quantizations, mainly owing to the smaller-dimension Hamiltonian matrices\cite{4SciRep4;7509}. Through modulating the various stacking configurations/symmetries, the physical phenomena become very rich and unique.

Apparently, the zone-folding effects thoroughly disappear in the sliding bilayer systems, clearly illustrating that their two pairs of ${2p_z}$-orbital-induced valence and conduction bands come to exist in the almost identical energy ranges. The low-energy essential properties are dominated by the low-lying electronic structures closely related to the K-/K$^\prime$-point valleys, regardless of the higher/deeper M-point valleys. The weak band overlaps clearly indicate the semi-metallic behaviors in all the sliding bilayer graphene systems. On the other hand, the $\theta$-dependent bilayer materials are zero-gap semiconductors with the monolayer-like [doubly degenerate] Dirac-cone structures, and the obvious saddle points might appear at low energies for the smaller-$\theta$ ones [e.g., (8,9) system in Fig. 2]. The relative shifts along the armchair direction can also create the dramatic transformations among the regular energy spectra: two vertical Dirac cones in AA stacking, two pairs of parabolic energy dispersions in AB configuration, and the non-vertical $\&$ tilted Dirac cones of AA$^\prime$. That is to say, the linear $\&$ isotropic Dirac cones are thoroughly destroyed, and then the other irregular/normal energy bands are formed as the intermediate electronic structures. The main features of zero-field band structures cover the serious distortions/hybridizations in energy dispersions/bands, the creation of arc-shaped stateless regions, more band-edge states near the K/K$^\prime$ points [saddle and extreme points], the strong dependence of band overlap on  the stacking configuration, the finite densities of states at the Fermi level [semimetals], and the gate-voltage-enriched characteristics,

Concerning the magnetic quantization phenomena, the sliding bilayer graphene systems exhibit two groups of ${2p_z}$-orbital-crated and conduction Landau levels and the shift-induced three kinds of Landau levels. The magneto-electronic states are directly magnetically quantized from two pairs of energy bands based on the tight-binding functions of four sublattices [details in Ref.\cite{4SciRep4;7509}]. The significant oscillation behaviors are characterized by the zero-point number of the localized subenvelope functions on the dominant sublattice. Obviously, the rich Landau levels can be classified into the well-e, perturbed, and undefined modes, in which they, respectively, possess a specific zero point, a major and some minor zero points, and the magnetic-field-dependent ones. The second kind, closely related to the non-monotonous energy dispersions/the non-single constant-energy loops  is accompanied by certain anti-crossings. Furthermore, the third kind, which mainly arises from the thorough destruction of Dirac-cone structures, lead to any anti-crossings between the first and second groups of conduction/valence Landau levels. Such a phenomenon could not survive under the twisted cases. As for the magneto-optical absorption spectra, the first, second and third kinds of Landau levels generate the ${\Delta\,n=\pm\,1}$ magneto-optical selection rules, the specific $\&$ extra ones, and the random vertical transitions with a lot of weaker absorption peaks. However, there are more Landau-level subgroups due to the specific Moire zone folding. As a result, the state degeneracy might be higher under the normal sliding, compared with the case of angle twisting. The magneto-optical properties and quantum Hall conductivities  are worthy of the further systematic studies using the combined generalized tight-binding model and Kubo formulas. In short, each stacking configuration in sliding bilayer materials presents the independent and unusual phenomenon, while that of twisted systems shows the hybridized characteristics associated with those in monolayer, and AA $\&$ AB stackings [the complex relations among the A${^l_i}$ and ${B^l_i}$ sublattices on the same and different graphene layers].

\par\noindent {\bf Acknowledgments}

This work was supported in part by the National Science Council of Taiwan,
the Republic of China, under Grant Nos. NSC 98-2112-M-006-013-MY4 and NSC 99-2112-M-165-001-MY3.

\newpage
\renewcommand{\baselinestretch}{0.2}

\newpage
\begin{figure}
\centering \includegraphics[width=0.9\linewidth]{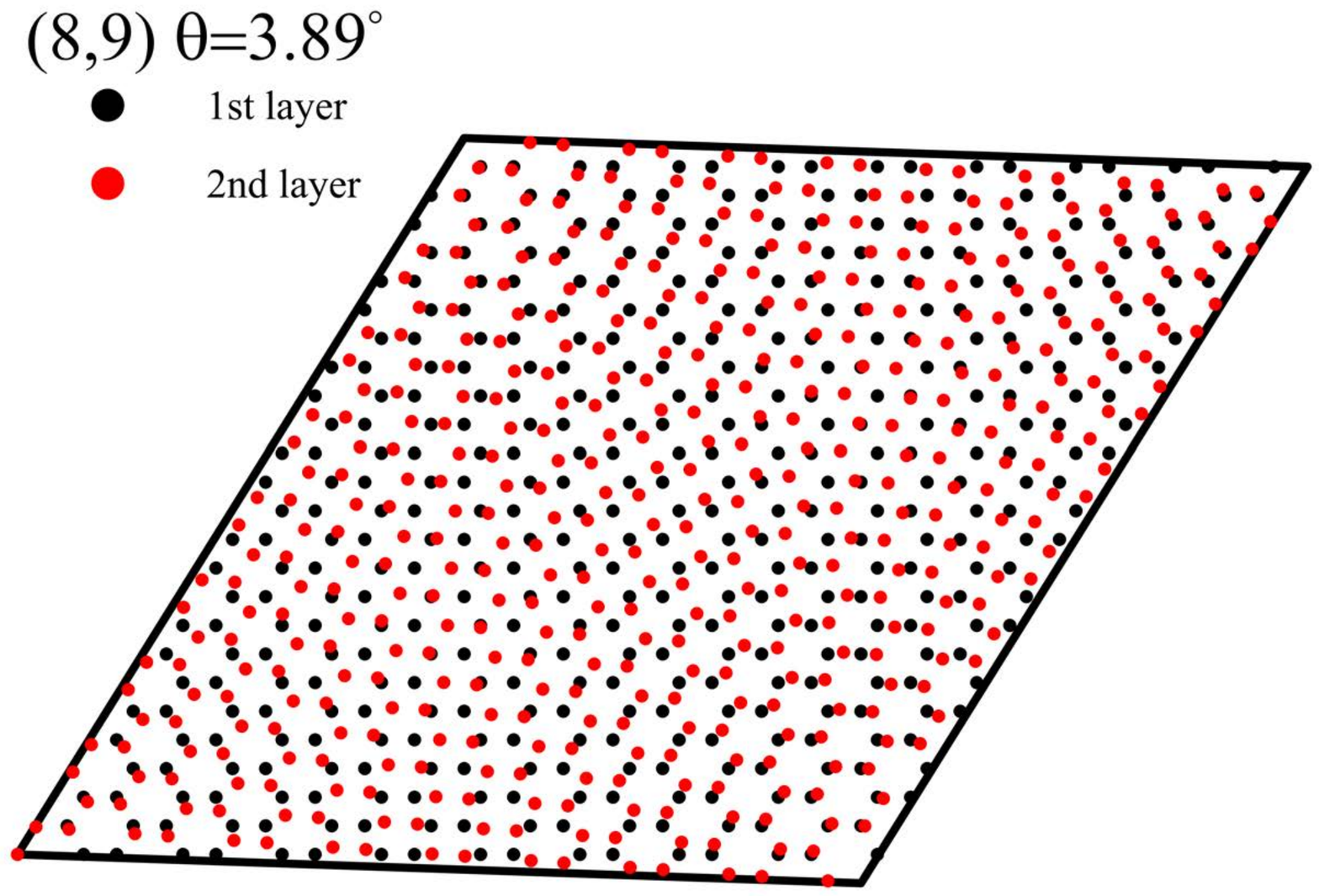}
\begin{center} Figure 1: The geometric structure of a twisted bilayer graphene, such as for a (1,2) system with fourteen carbon atoms in a unit  cell for each layer.
\end{center} \end{figure}
\newpage

\begin{figure}
\centering \includegraphics[width=0.9\linewidth]{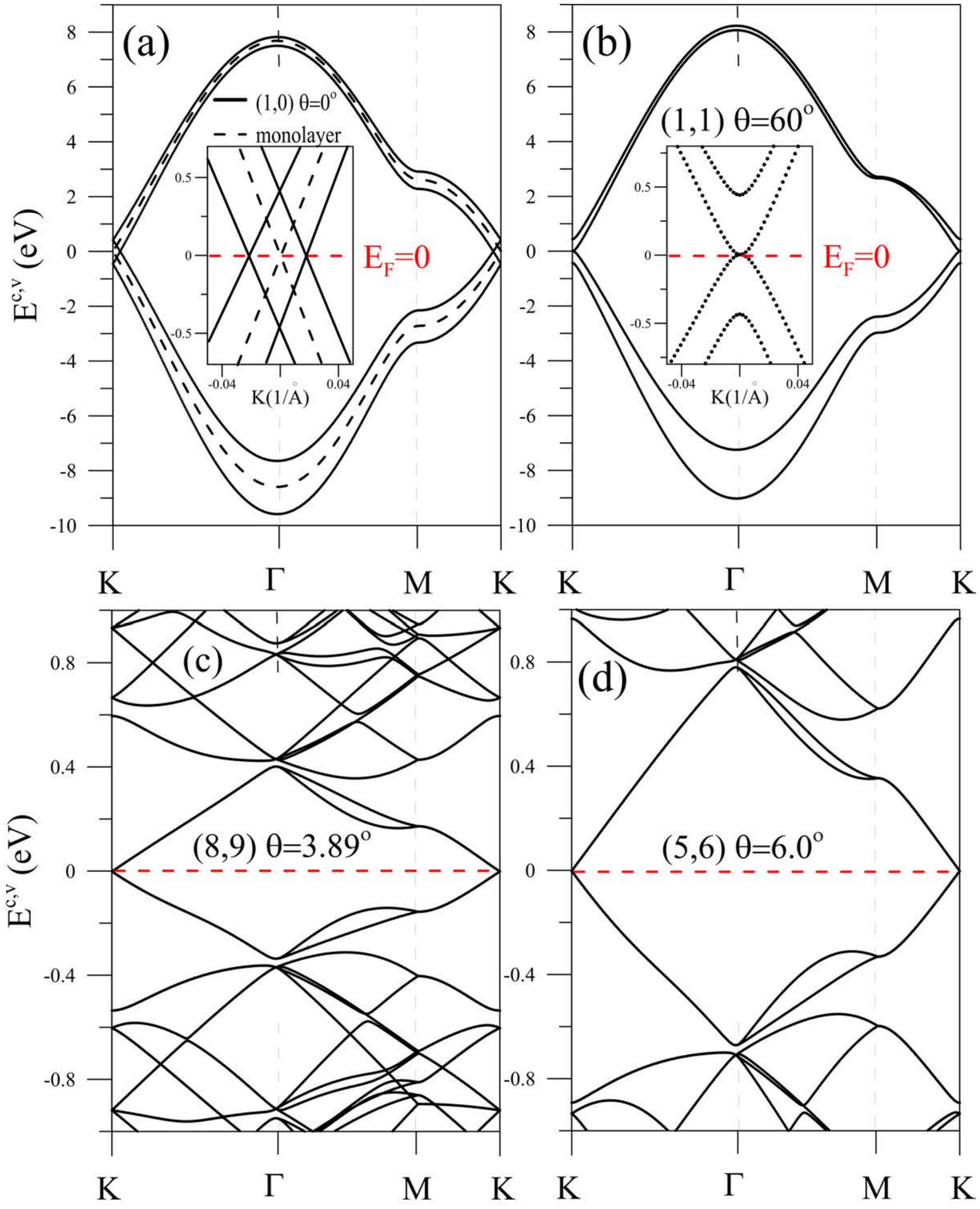}
\begin{center} Figure 2: The band structures due to the C-${2p_z}$ orbitals for bilayer graphene systems with the different twisted angles: (a) ${\theta\,=0^\circ}$ [AA], (b) 60$^\circ$ [AB], (c) 3.89$^\circ$ [(8,9)] and (d) 6.0$^\circ$ [(5,6)]. Also shown in (a) is that of monolayer graphene by the dashed curve.
\end{center} \end{figure}
\newpage
\begin{figure}
\centering \includegraphics[width=0.9\linewidth]{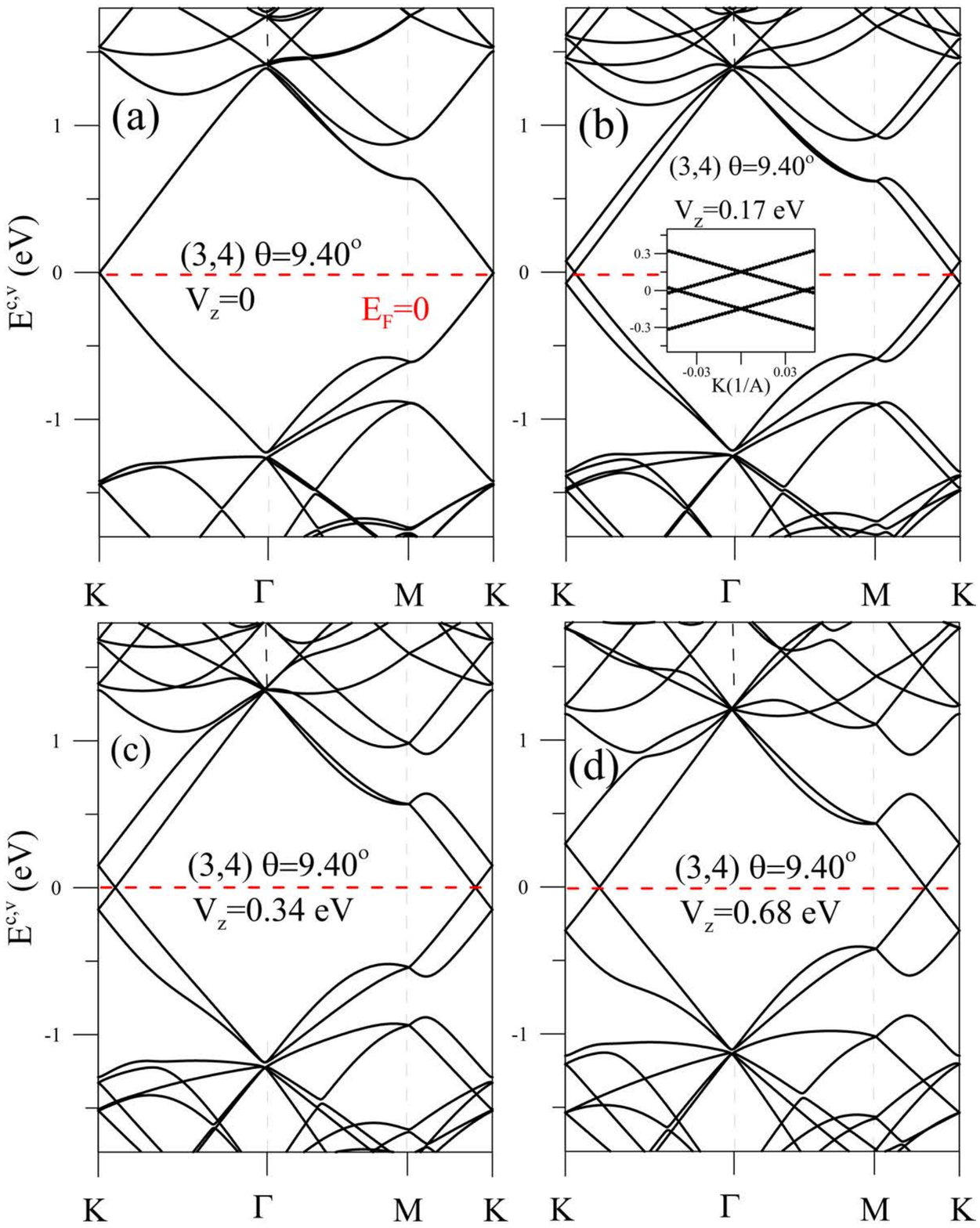}
\begin{center} Figure 3: The band structures of the twisted (3,4) bilayer graphene [${\theta\,=9.40^\circ}$] under the various gate voltages: (a) ${V_z=0}$, (b) 0.17 eV, (c) 0.34 eV and (d) 0.68 eV.
\end{center} \end{figure}
\newpage
\begin{figure}
\centering \includegraphics[width=0.9\linewidth]{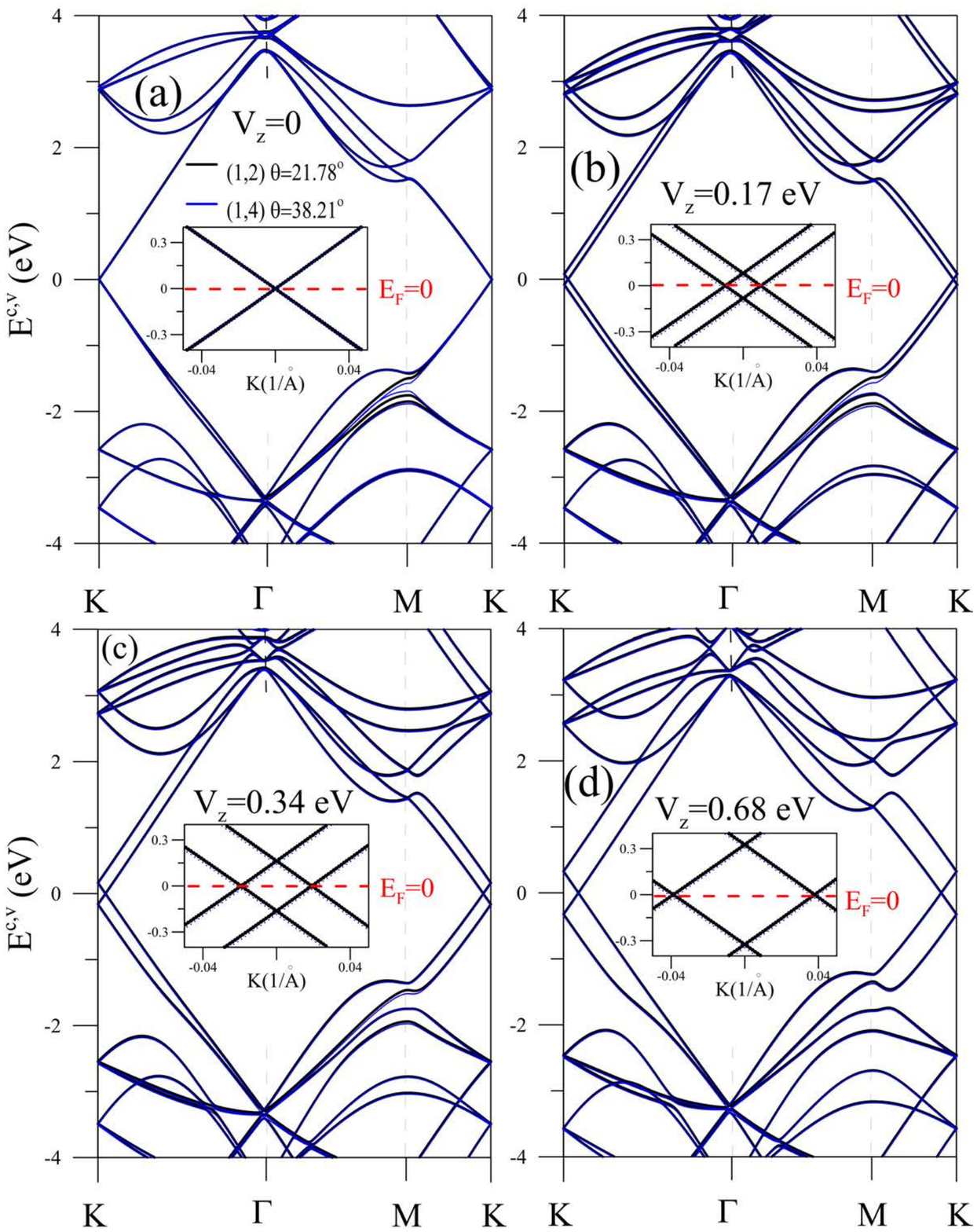}
\begin{center} Figure 4: Similar plot as Fig. 4.3, being shown for the (1,2) bilayer graphene with ${\theta\,=21.78^\circ}$ at different gate voltages: (a) ${V_z=0}$, (b) 0.17 eV, (c) 0.37 eV and (d) 0.68 eV. Also for comparison is that of ${(1,4)}$ system [${\theta\,=38.21^\circ}$; ${V_z=0}$ by the blue curves].
\end{center} \end{figure}
\newpage
\begin{figure}
\centering \includegraphics[width=0.9\linewidth]{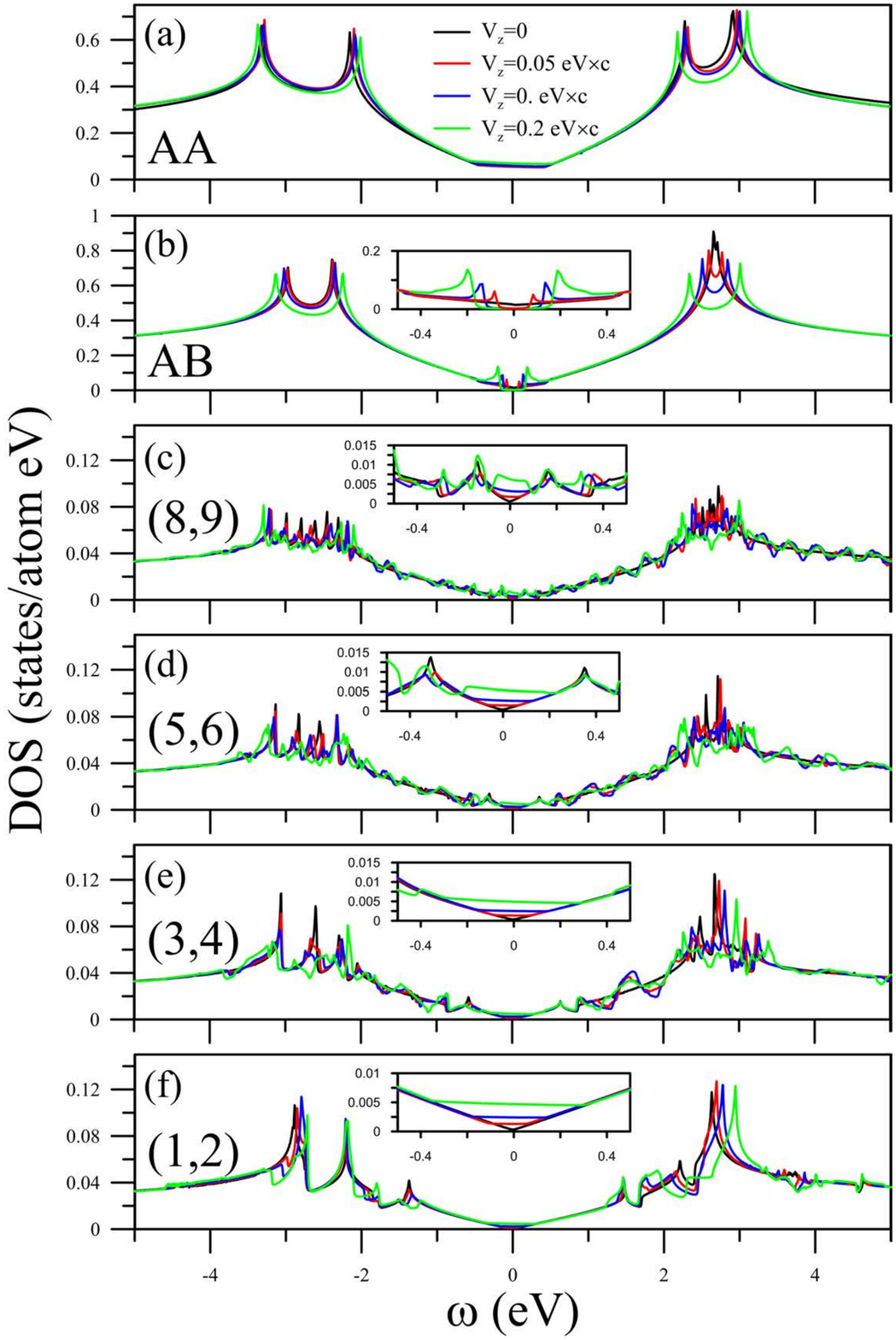}
\begin{center} Figure 5. the twisted-angle- and gate-voltage-dependent densities of states for the bilayer graphene systems under ${V_z}$=0. 0.05 eV, 0.10 eV and 0.20 eV: (a) AA, (b) AB, (c) (8,9), (d), (5,6), (e) (3,4) and (f) (1,2).
\end{center} \end{figure}
\newpage
\begin{figure}
\centering \includegraphics[width=0.9\linewidth]{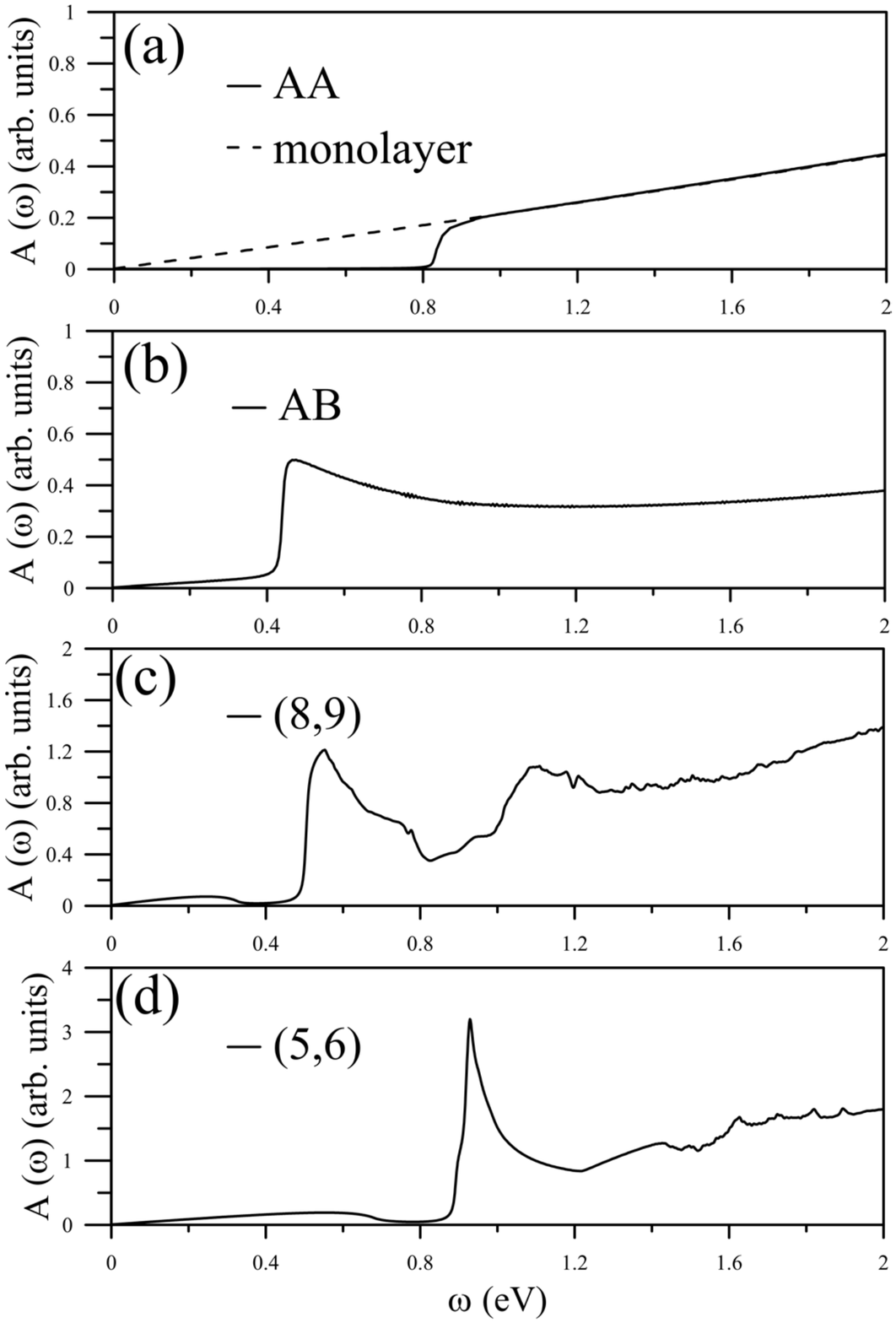}
\begin{center} Figure 6. The optical absorption spectra for (a) AA, (b) AB, (c) (8,9) and (d) (5,6). Also shown in (a) is that of monolayer graphene by the dashed curve.
\end{center} \end{figure}
\newpage
\begin{figure}
\centering \includegraphics[width=0.9\linewidth]{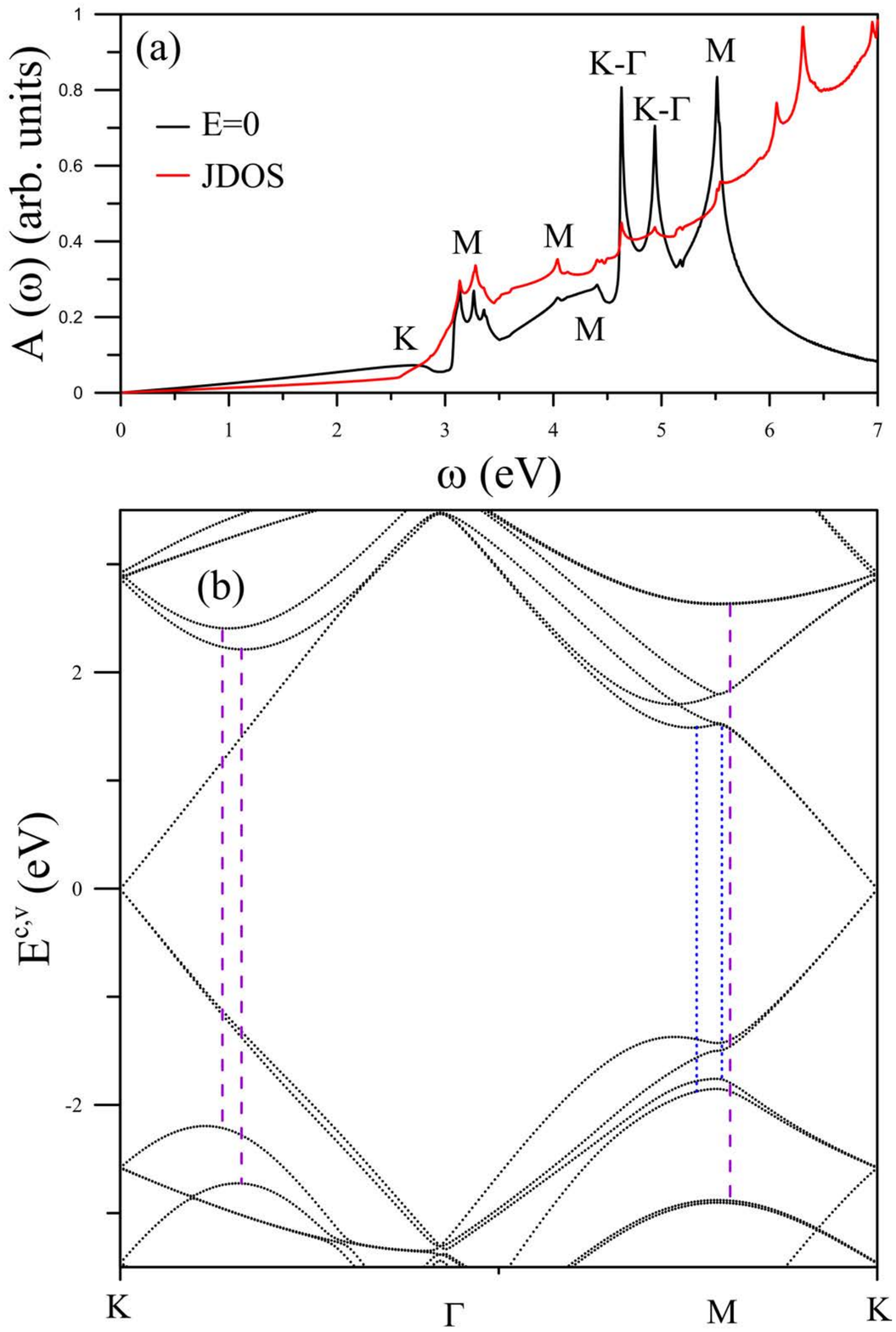}
\begin{center} Figure 7: (a) The joint densities of state [the red curve] and optical excitation [the black curve]
for the (1,2) bilayer graphene, and (b) the significant vertical transition channels corresponding to the strong absorption structures through the dashed blue and purple lines for the initial two and others, respectively.
\end{center} \end{figure}
\newpage
\begin{figure}
\centering \includegraphics[width=0.9\linewidth]{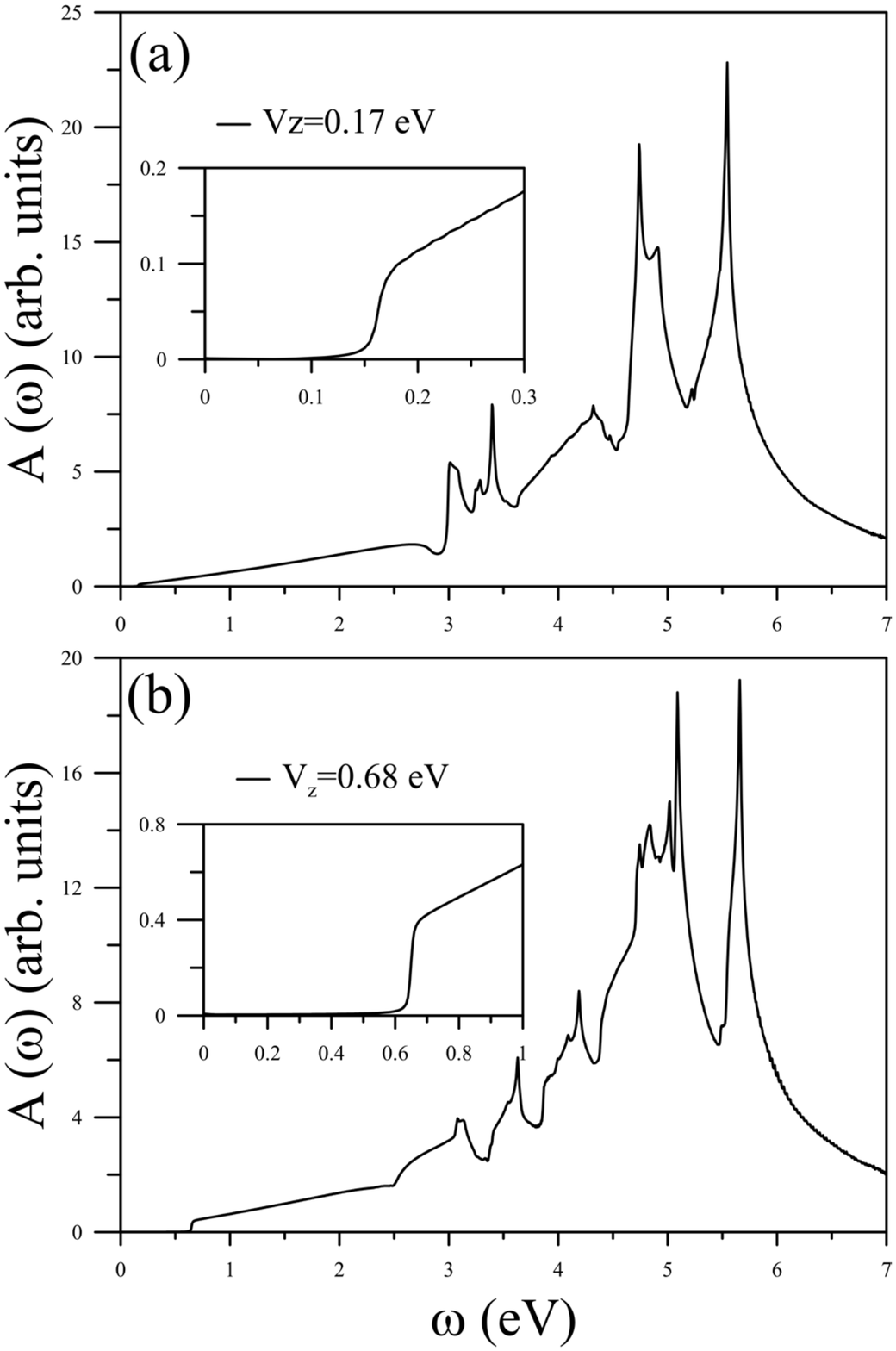}
\begin{center} Figure 8(a)(b): The gate-voltage-enriched absorption spectra of the (1,2) bilayer graphene at (a) ${V_z=0.17}$ eV, (b) ${V_z=0.68}$ eV and (c) ${V_z=0.34}$ eV; furthermore, (d) certain vertical transitions under the (c) case.
\end{center} \end{figure}
\newpage
\begin{figure}
\centering \includegraphics[width=0.9\linewidth]{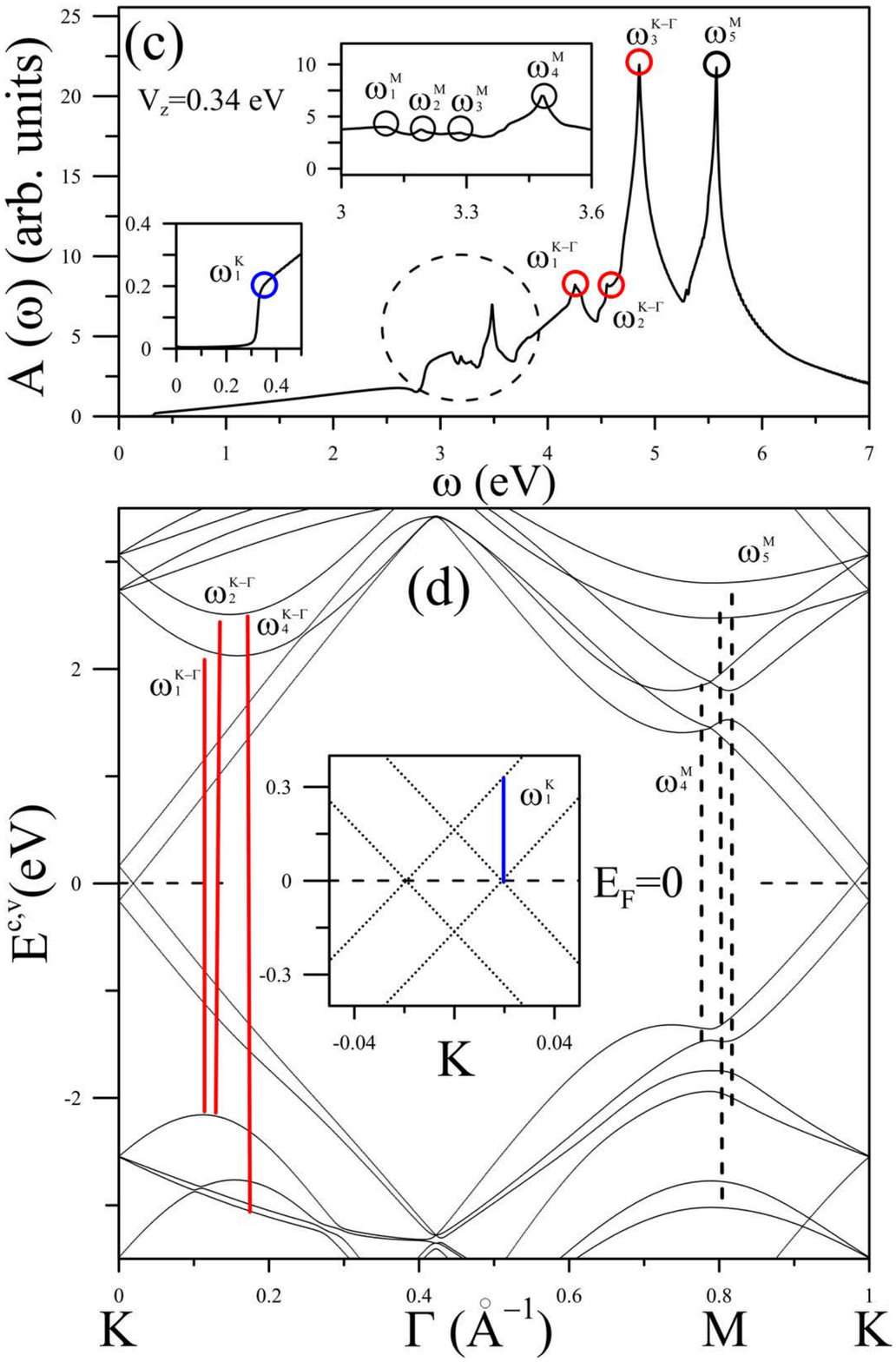}
\begin{center} Figure 8(c)(d): The gate-voltage-enriched absorption spectra of the (1,2) bilayer graphene at (a) ${V_z=0.17}$ eV, (b) ${V_z=0.68}$ eV and (c) ${V_z=0.34}$ eV; furthermore, (d) certain vertical transitions under the (c) case.
\end{center} \end{figure}
\newpage
\begin{figure}
\centering \includegraphics[width=0.9\linewidth]{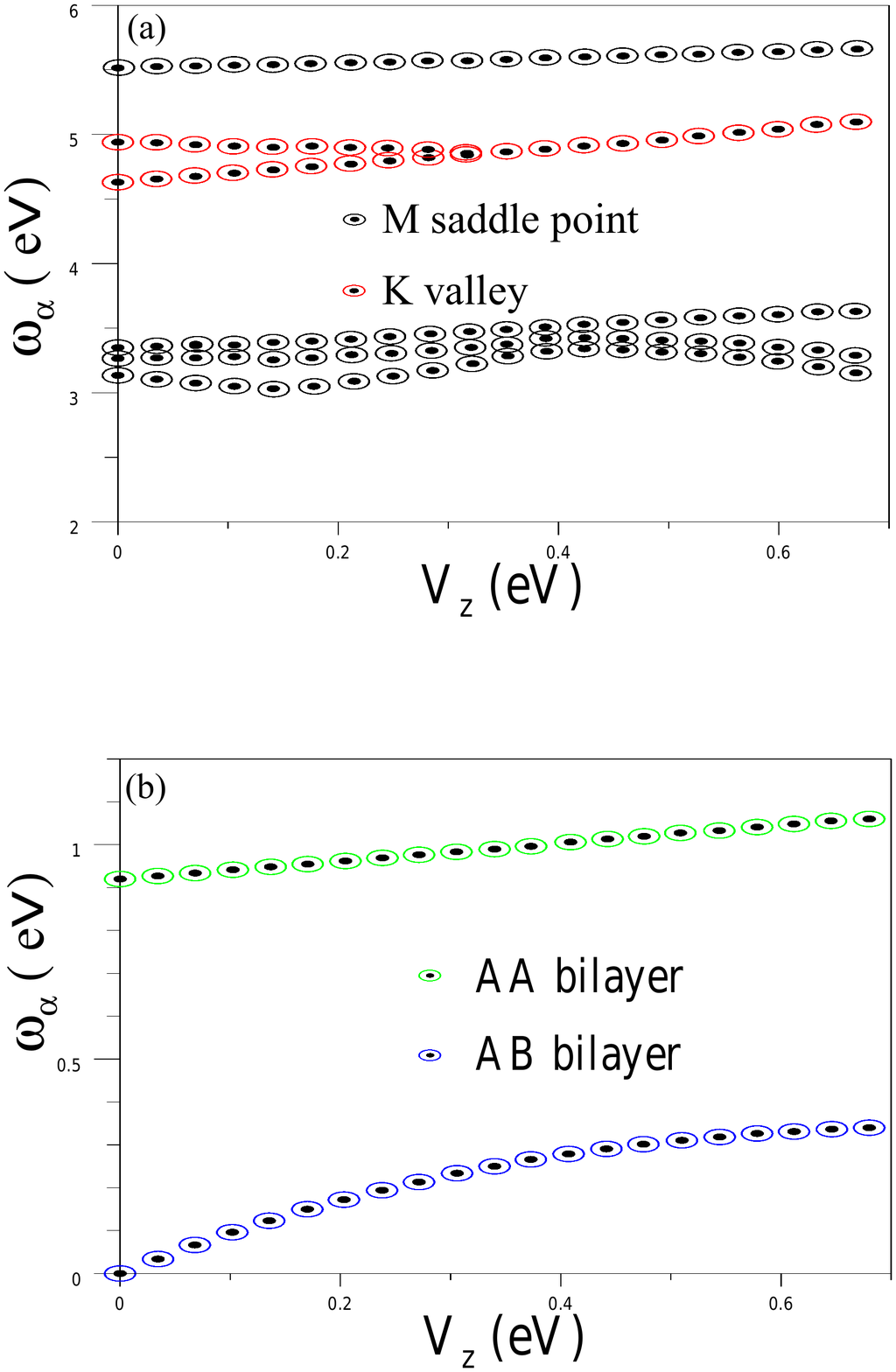}
\begin{center} Figure 9: The gate-voltage-dependent excitation frequencies corresponding to (a) the low-lying K valley and M saddle structure in the bilayer (1,2) graphene system and (b) the optical gaps of the AA and AB stackings.
\end{center} \end{figure}
\newpage
\begin{figure}
\centering \includegraphics[width=0.9\linewidth]{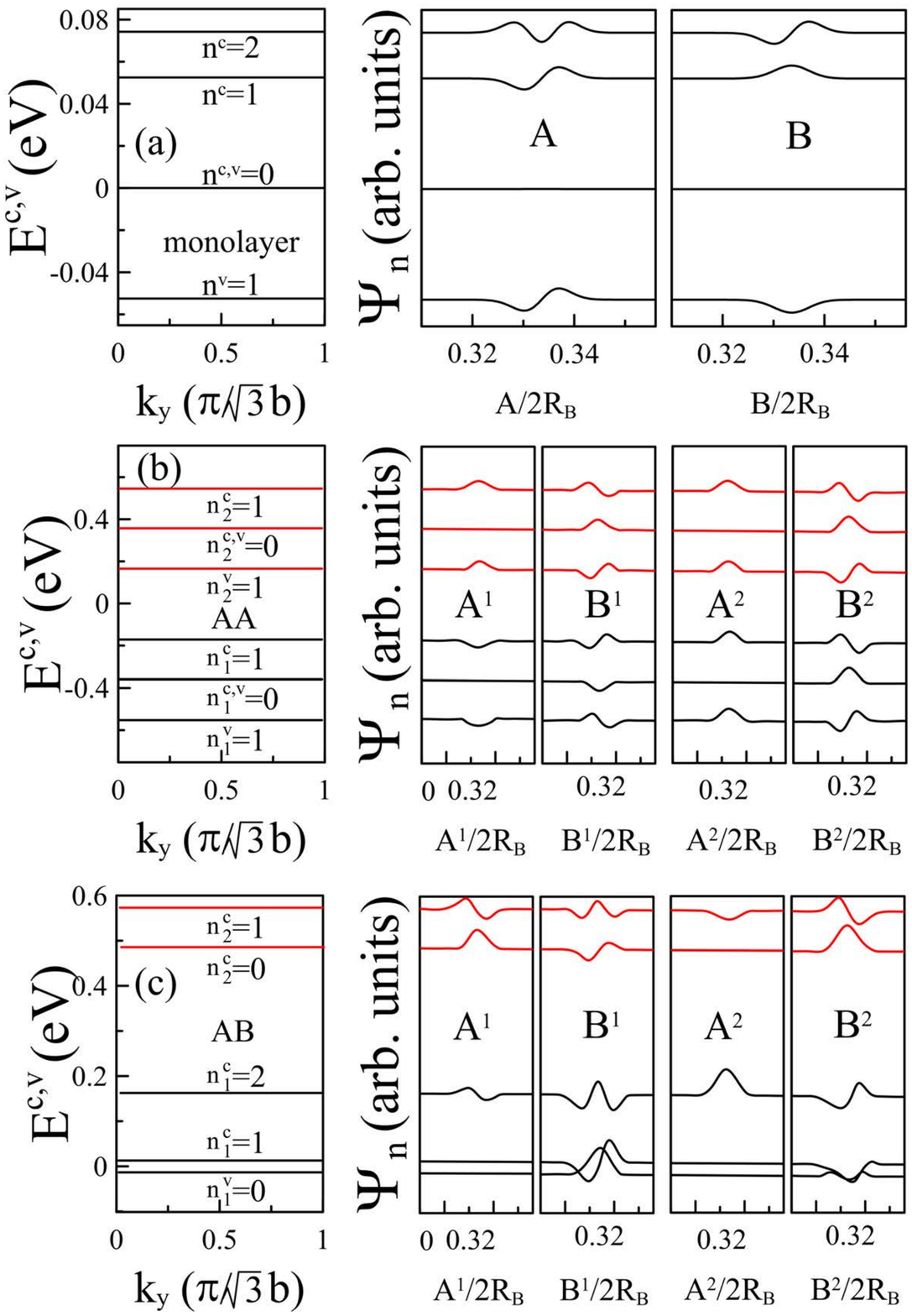}
\begin{center} Figure 10: The Landau-level energies and subenvelope functions for the (a) monolayer and bilayer (b) AA $\&$ (c) AB staclings.
\end{center} \end{figure}
\newpage
\begin{figure}
\centering \includegraphics[width=0.8\linewidth]{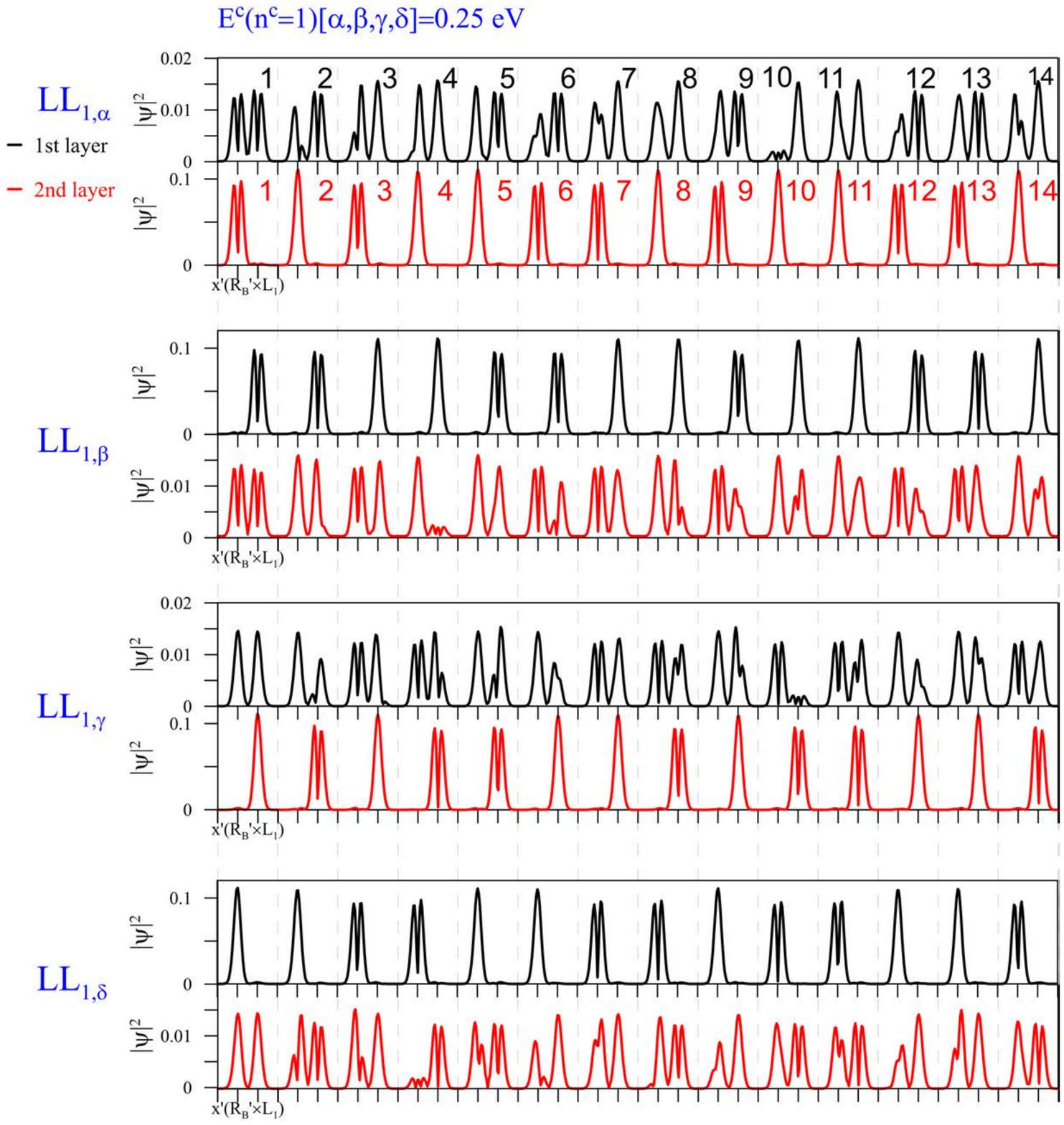}
\begin{center} Figure 11: The magneto-electronic wave functions of the twisted (1,2) bilayer graphene, being built from the amplitudes of the  14-sublattices tight-binding functions on the upper and lower layers [the black and red curves], for the second conduction Landau levels of the first and second subgroups with the four degenerate states $\alpha$, $\beta$, $\gamma$ and $\delta$.
\end{center} \end{figure}
\newpage
\begin{figure}
\centering \includegraphics[width=0.9\linewidth]{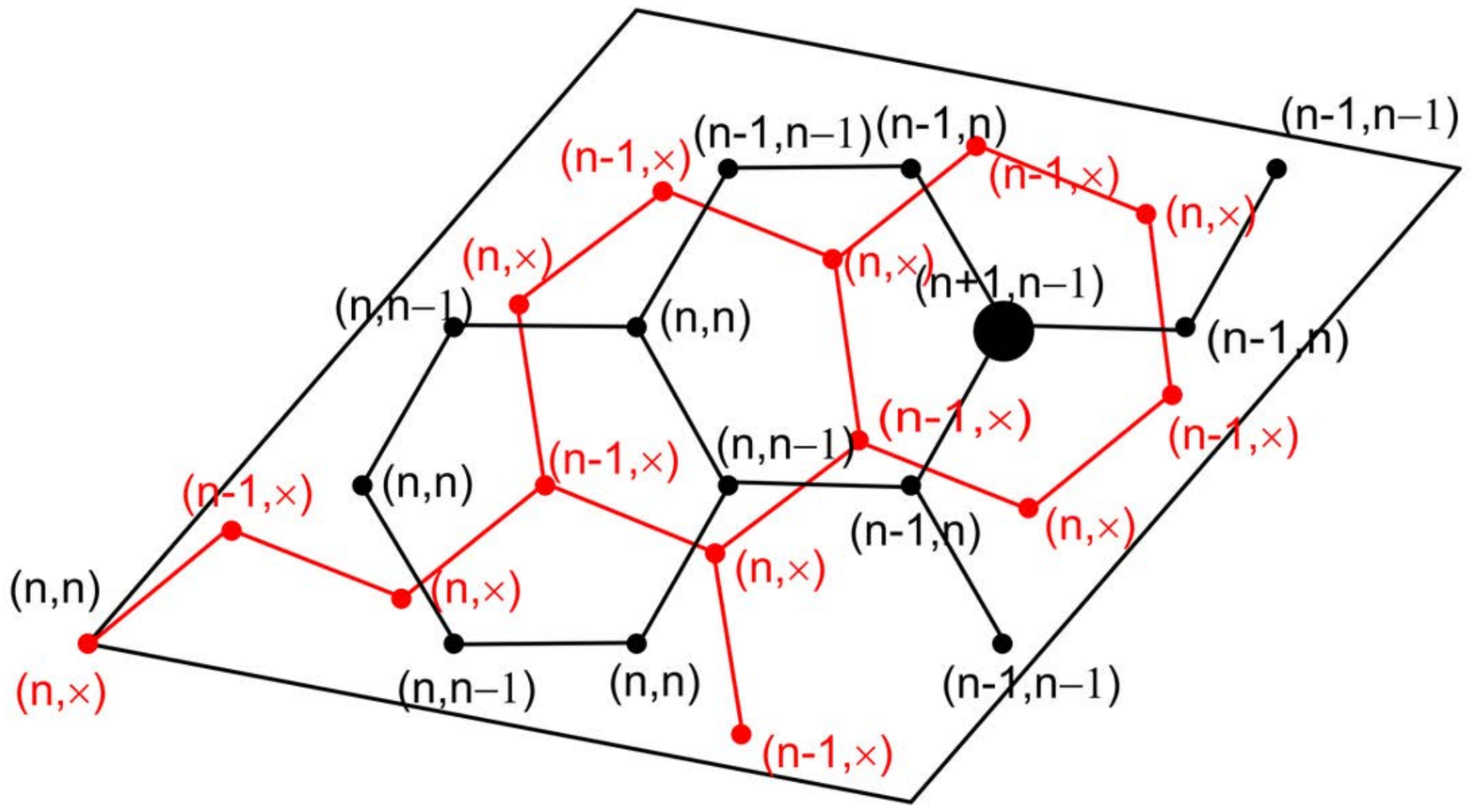}
\begin{center} Figure 12(a): For the (1,2) bilayer graphene, the oscillation modes of the fourteen sublattices in the upper and lower layers, being illustrated for the $\alpha$ state at ${B_z=100}$T and (a) ${V_z=0}$ and (b) ${V_z=0.34}$ eV.
\end{center} \end{figure}
\newpage
\begin{figure}
\centering \includegraphics[width=0.9\linewidth]{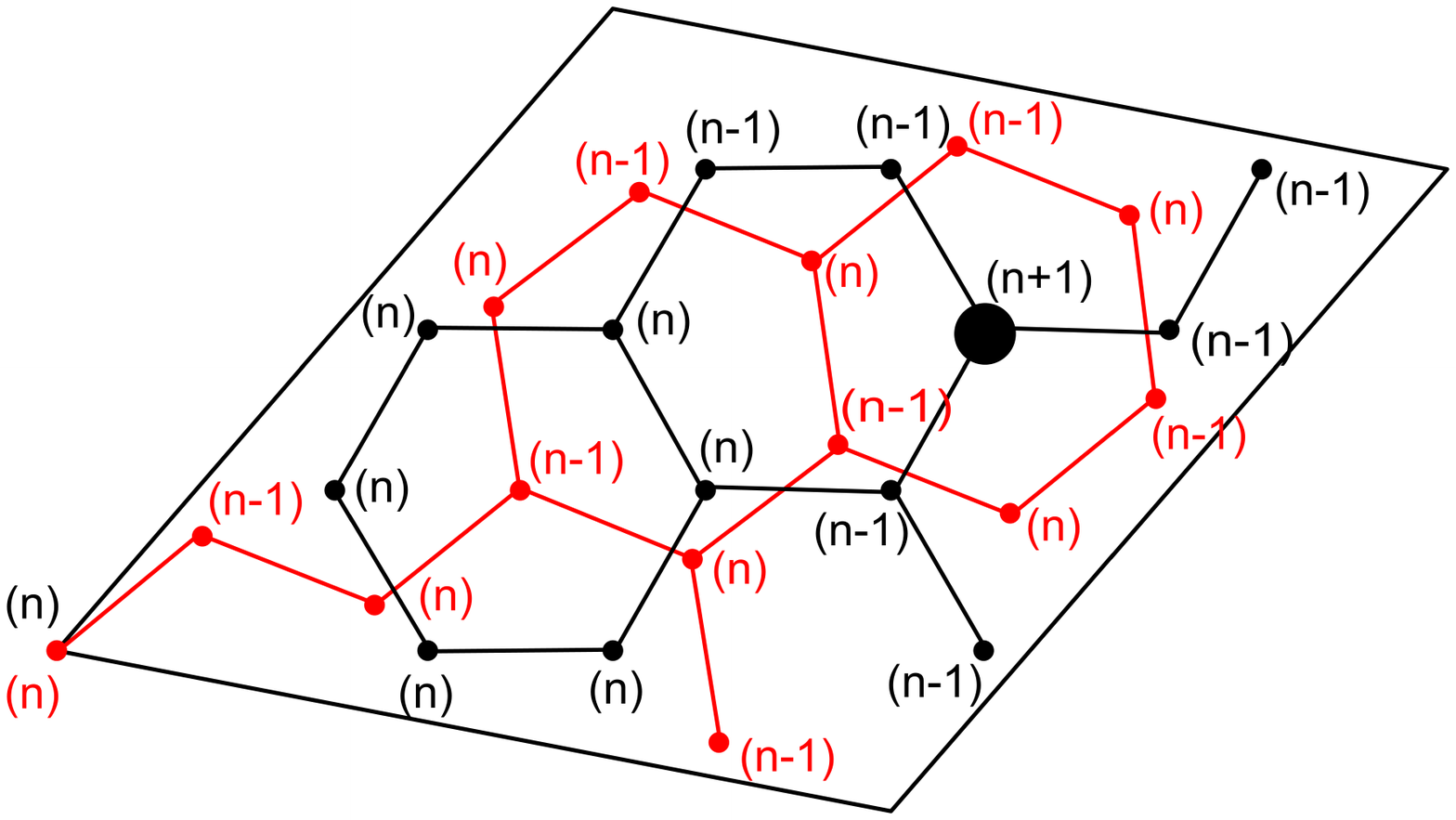}
\begin{center} Figure 12(b): For the (1,2) bilayer graphene, the oscillation modes of the fourteen sublattices in the upper and lower layers, being illustrated for the $\alpha$ state at ${B_z=100}$T and (a) ${V_z=0}$ and (b) ${V_z=0.34}$ eV.
\end{center} \end{figure}
\newpage
\begin{figure}
\centering \includegraphics[width=0.9\linewidth]{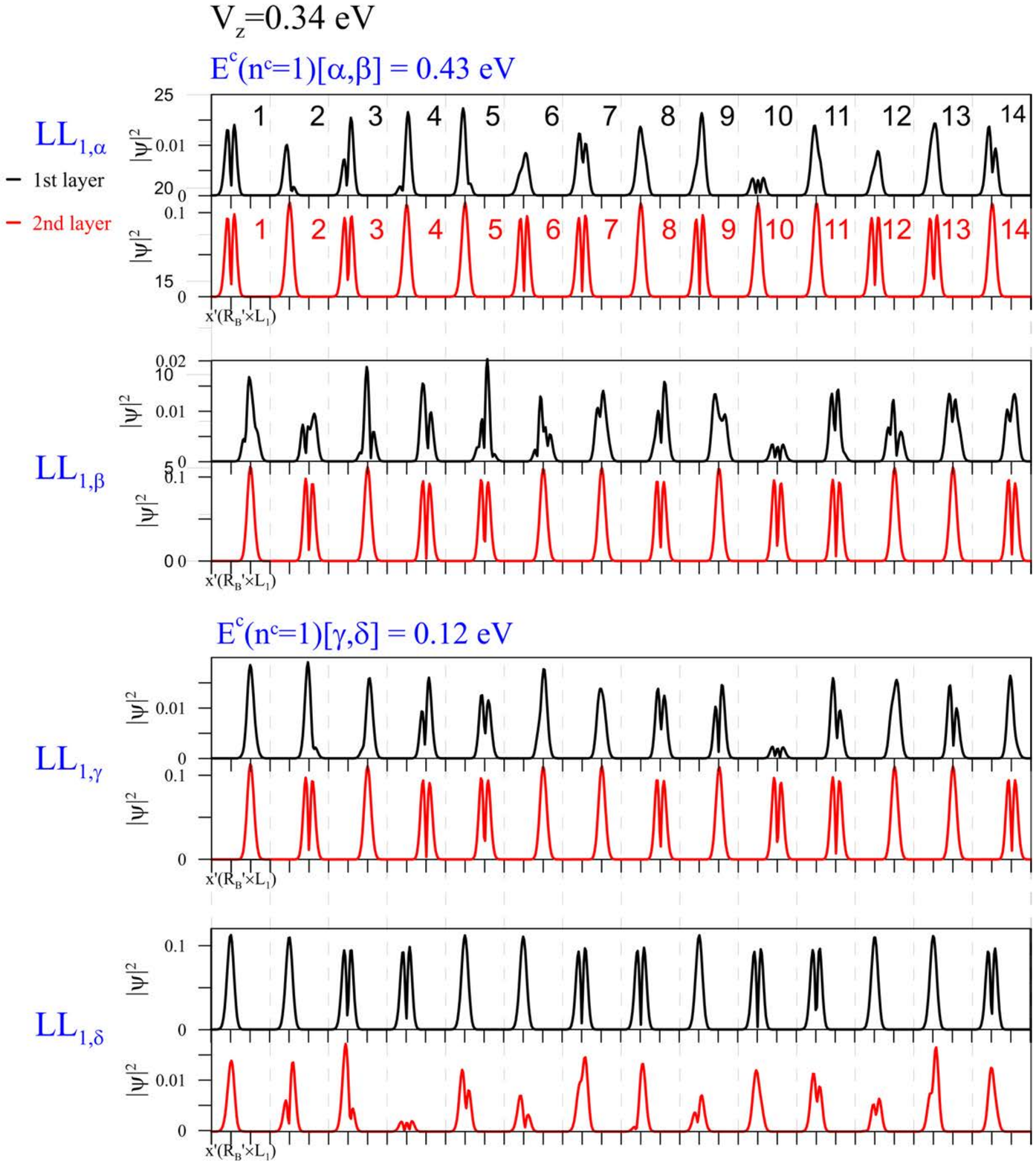}
\begin{center} Figure 13: Similar plot as Fig. 11, while displayed under ${V_z=0.34}$ eV.
\end{center} \end{figure}
\newpage
\begin{figure}
\centering \includegraphics[width=0.9\linewidth]{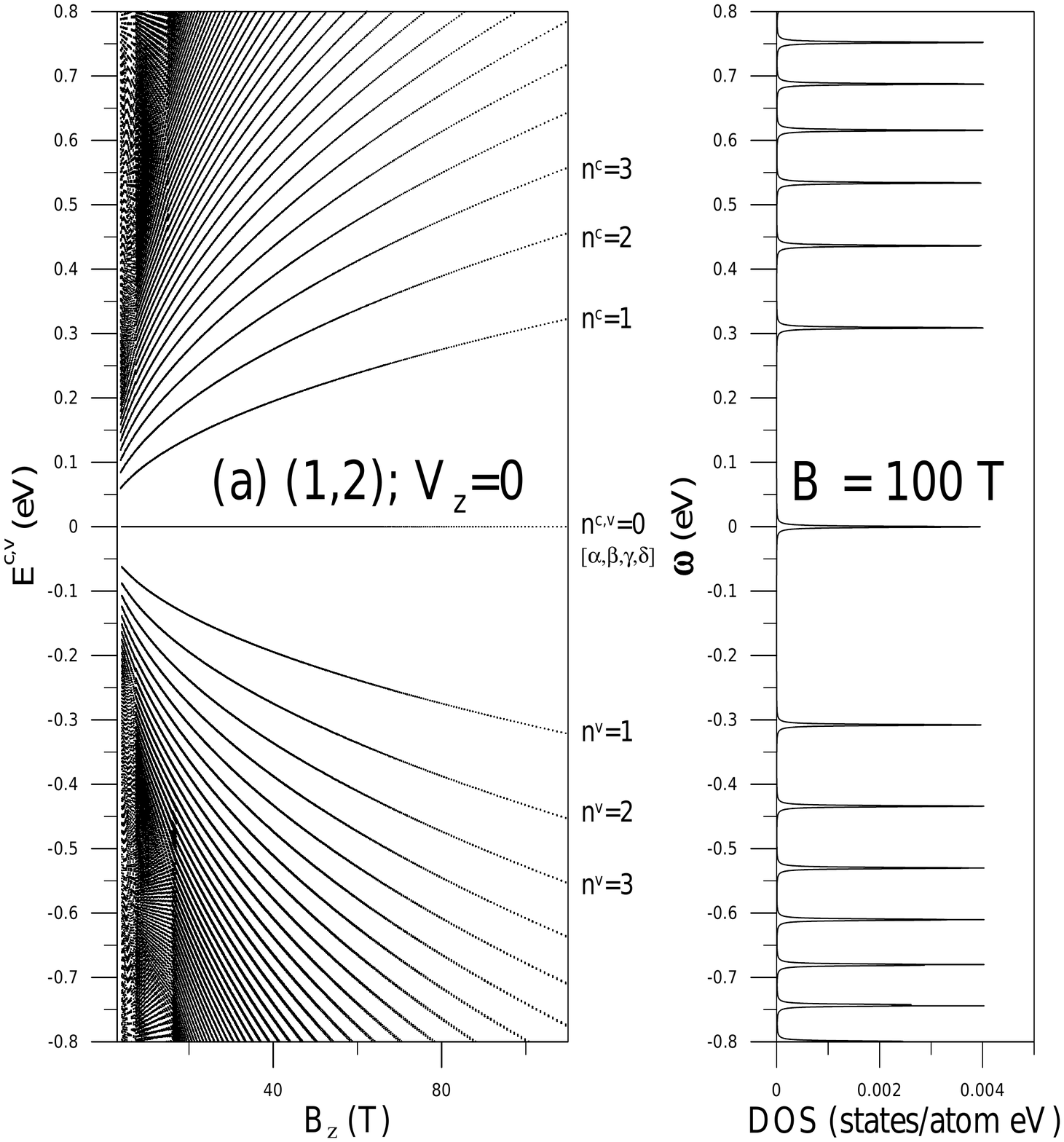}
\begin{center} Figure 14: (a) The magnetic-field-dependent Landau-level energy spectrum of the (1,2) bilayer graphene system, being accompanied with (b)the density of states.
\end{center} \end{figure}
\newpage
\begin{figure}
\centering \includegraphics[width=0.9\linewidth]{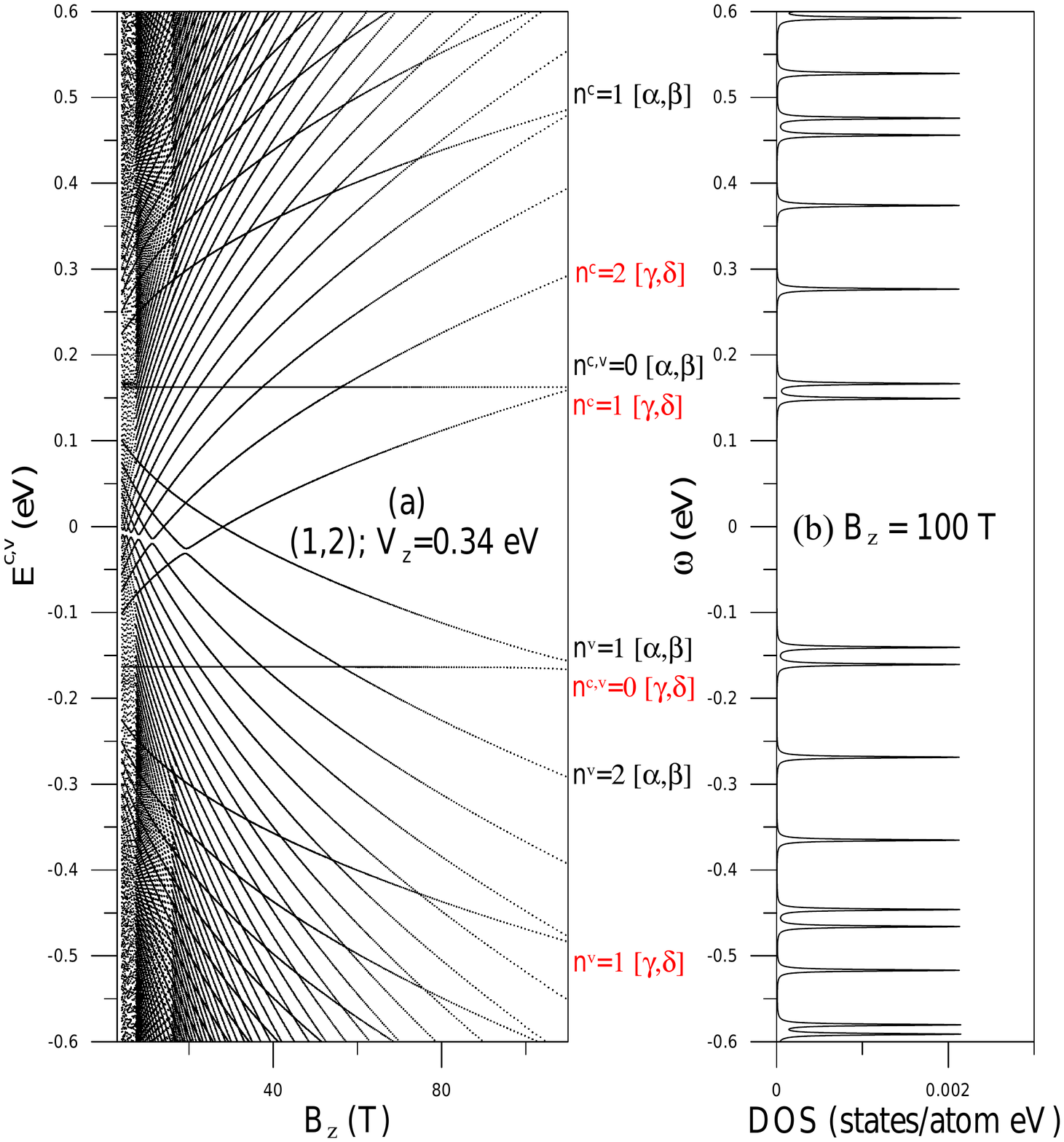}
\begin{center} Figure 15: Similar plot as Fig. 14, but shown at (a) ${V_z=0.34}$ eV with (b) the density of states.
\end{center} \end{figure}
\newpage
\begin{figure}
\centering \includegraphics[width=0.9\linewidth]{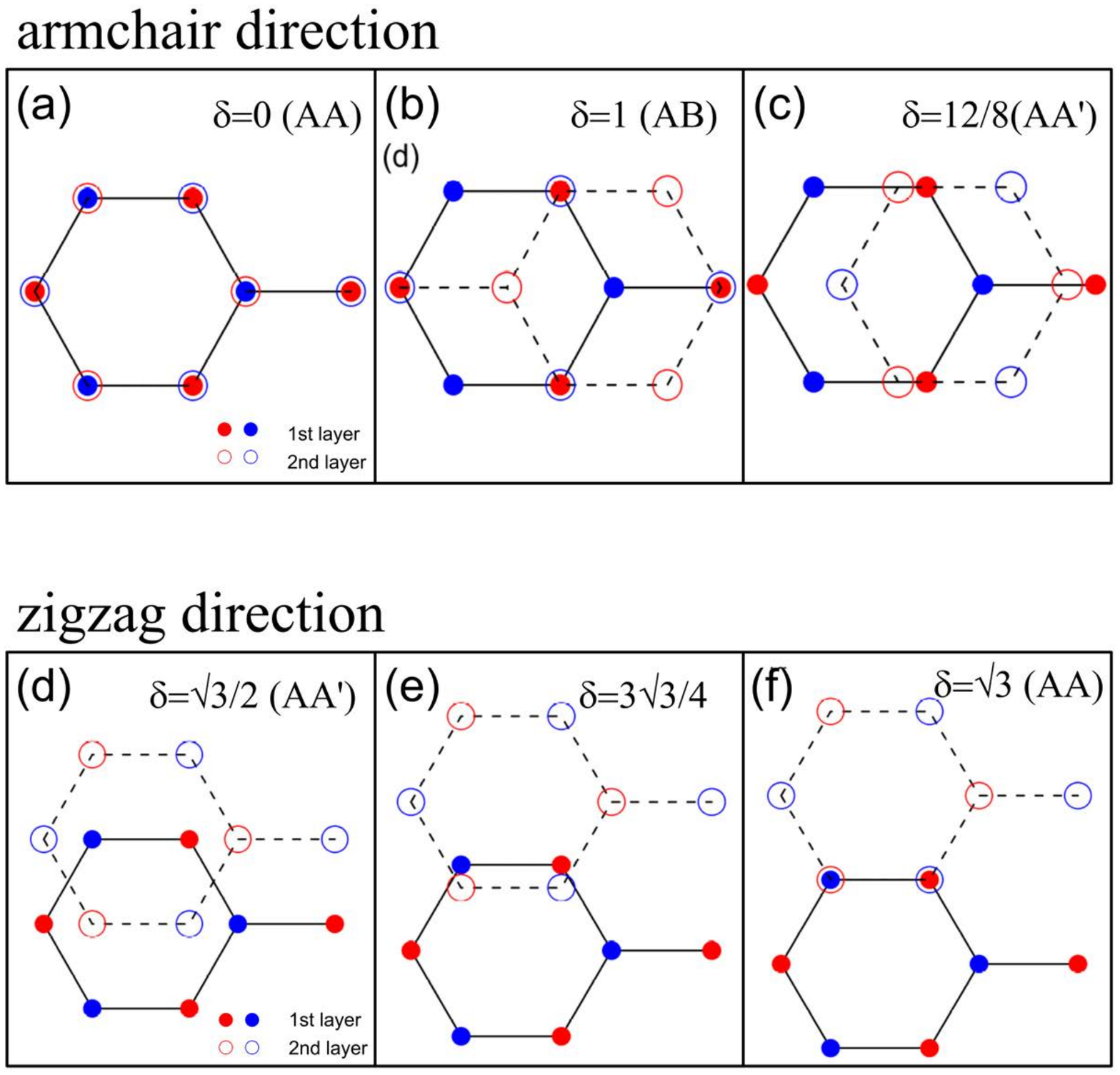}
\begin{center} Figure 16: The sliding bilayer graphene systems with the relative shifts along the armchair and zigzag directions: AA${\rightarrow}$AB${\rightarrow}$AA$^\prime$ \& AA$^\prime$${\rightarrow}$AA.
\end{center} \end{figure}

\end{document}